%% file: main.tex
\documentclass[sigconf,authorversion,nonacm,balance=false]{acmart}
\usepackage{popets}

\setcopyright{popets}
\copyrightyear{2024}

\acmYear{2024}
\acmVolume{2024}
\acmNumber{v1}
\acmDOI{XXXXXXX.XXXXXXX}
\acmISBN{}
\acmConference{Arxiv Version}
\usepackage{graphicx}
\usepackage{subfigure}
\usepackage{booktabs}
\usepackage{multirow}
\def\until{\mathbf{U}}
\usepackage{algcompatible}
\usepackage[ruled,vlined,linesnumbered]{algorithm2e}
\usepackage{empheq}
\usepackage{xcolor}
\usepackage{rotating}
\usepackage{siunitx}

\newcommand{\Autoref}[1]{%
  \begingroup%
  \def\chapterautorefname{Chapter}%
  \def\sectionautorefname{Section}%
  \def\subsectionautorefname{Section}%
  \def\algorithmautorefname{Algorithm}%
  \autoref{#1}%
  \endgroup%
}

\newcommand{\ruletemplate}[1]{{\sf \small #1}}

\newcommand\shortsection[1]{\paragraph{\noindent\bf #1.}}

\SetCommentSty{mycommfont}

\begin{document}

\title[DP-RuL]{DP-RuL: Differentially-Private Rule Learning for Clinical Decision Support Systems}


\author{Josephine Lamp}
\affiliation{%
  \institution{University of Virginia}
  \city{}
  \state{}
  \country{}}
\email{}

\author{Lu Feng}
\affiliation{%
  \institution{University of Virginia}
  \city{}
  \state{}
  \country{}}
\email{}

\author{David Evans}
\affiliation{%
  \institution{University of Virginia}
  \city{}
  \state{}
  \country{}}
\email{}


\renewcommand{\shortauthors}{Lamp et al.}

\begin{abstract}
Serious privacy concerns arise with the use of patient data in rule-based clinical decision support systems (CDSS).  
The goal of a privacy-preserving CDSS is to learn a population ruleset from individual clients' local rulesets, while protecting the potentially sensitive information contained in the rulesets. 
We present the first work focused on this problem and develop a framework for learning population rulesets with local differential privacy (LDP), suitable for use within a distributed CDSS and other distributed settings. 
Our rule discovery protocol uses a Monte-Carlo Tree Search (MCTS) method integrated with LDP to search a rule grammar in a structured way and find rule structures clients are likely to have. Randomized response queries are sent to clients to determine promising paths to search within the rule grammar. 
In addition, we introduce an adaptive budget allocation method which dynamically determines how much privacy loss budget to use at each query, resulting in better privacy-utility trade-offs. We evaluate our approach using three clinical datasets and find that we are able to learn population rulesets with high 
coverage (breadth of rules) and 
clinical utility even at low privacy loss budgets.
\end{abstract}

\keywords{clinical decision support systems, local differential privacy, monte-carlo tree search, signal temporal logic}

\maketitle

\input{1.intro}

\input{2.prelim}
\input{3.approach}

\input{4.experiments}
\input{5.rel-work}

\input{6.conclusion}


\bibliographystyle{ACM-Reference-Format}
\bibliography{refs}

\input{appendix}

\end{document}

%% file: 1.intro.tex
\section{Introduction}\label{sec:intro}

\begin{figure}[t]
\centering
\includegraphics[width=0.55\linewidth]{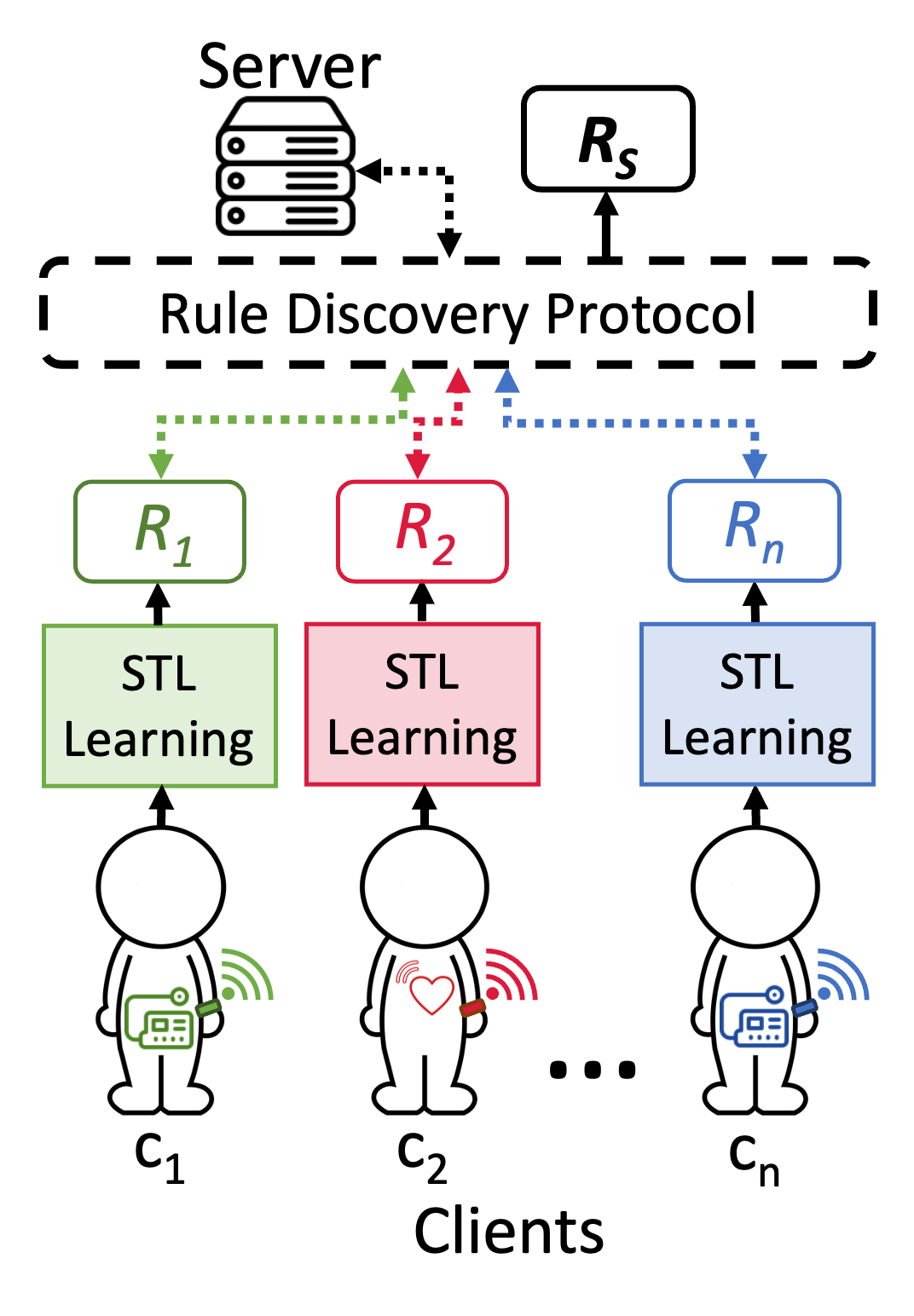}
\caption{Our privacy-preserving CDSS framework. Clients locally collect data from sensors and wearables, which are used to learn personalized rule sets ($R_1, \ldots, R_n$) using STL Learning describing individual conditions. A \emph{Rule Discovery Protocol} sends a series of structured queries to the clients who respond using randomized response, to produce an aggregate population ruleset $R_S$ to discover generalizable clinical rules.}
\Description{}
\label{fig-examp-cdss-arch}
\end{figure}

A Clinical Decision Support System (CDSS) provides information
to aid health care providers and patients in making clinical decisions~\cite{musen2014clinical}. 
With the availability of mobile sensors and devices, CDSSs 
are being integrated into third-party health applications for a myriad of health contexts, including chronic disease management, remote patient monitoring, and medical triage~\cite{martinez2014mobile}. 
CDSSs often use machine learning (ML) technologies to aggregate patient data. In particular, many CDSSs rely on logic-based learning systems~\cite{latif2020implementation}, in which structured \emph{rules} are used to make decisions due to their increased expressivity (diverse representations of medical associations), dual understandability by humans and machines (e.g., using a rule grammar), and increased explainability which promotes user trust in the system~\cite{watson2019clinical, gretton2018trust}. 
Even with the proliferation of deep learning and generative ML, rule-based learners are still extremely common in clinical applications; indeed, some deep learning frameworks actually use a rule-based output layer or ensemble learner to better 
explain model predictions, increasing trust and interpretability of 
the overall system~\cite{kierner2023taxonomy, berge2023combining}. 
In a typical distributed CDSS setting, mobile apps using data from wearables learn and characterize patient behaviors using a rule-based learner, such as Signal Temporal Logic (STL) Learning (described in \Autoref{sec-prelim-stl}).  From there, the apps send the rules to a centralized server which aggregates patterns across patients to learn about clinical conditions that may generalize to broader populations or subcohorts. 

Serious privacy concerns, such as data compromise and unsanctioned use of user data, arise with the use of patient data in CDSSs, especially those deployed in third-party health applications~\cite{lange2023privacy}. 
Since these third-party health applications are not covered by HIPAA, they are not subject to the same protective privacy requirements that govern data in health organizations~\cite{Price2019}.
Breaches of patient data from third party health apps, however, can have significant consequences, including job and insurance discrimination based on exposed sensitive health details (e.g., a patient's past drug, mental health or serious disease history)~\cite{insights2019}. 

\shortsection{Project Goal}
Given these concerns, the goal of this project is to learn a population ruleset representative of the local client rule structures, while preserving the privacy of individuals involved in the rule collection. 
We consider an untrusted server $S$ that wishes to generate a population ruleset $R_S$ from the local rulesets of $n$ individual clients, $\{R_1, ..., R_n\}$. Participating clients are expected to behave honestly but want to protect the sensitive information contained in their rulesets from the server and other protocol participants. 
We wish to learn population rulesets with two key qualities: (1) \textit{coverage} --- the learned population ruleset captures well the breadth of behavior of the client population; and (2) \textit{clinical utility} --- the learned rules are useful in a medical context. 
\shortsection{Learning Rules with Privacy}
To provide local differential privacy (LDP), individual users each perturb their own data before it is collected and used for population-level aggregation~\cite{kasiviswanathan2011can}. 
Previous work has developed differentially-private methods for distributed learning in various settings 
including finding new frequent strings~\cite{Fanti2015}, discovering keystroke data~\cite{kim2018learning, wang2018privtrie}, text mining~\cite{wang2020federated}, frequent item mining~\cite{wang2018locally,wang2019locally,jia2019calibrate} and data mining personal information~\cite{fletcher2019decision}. However, as we discuss more in \Autoref{sec:related}, no previous work has developed LDP methods for learning logic-based rule structures or for CDSS applications, and none of the methods developed for these other settings can be directly applied to provide an adequate solution to the privacy rule discovery problem. 
\shortsection{Contributions}
We present and evaluate the first LDP framework to learn population rulesets with high coverage and clinical utility for logic-based CDSSs, depicted in \Autoref{fig-examp-cdss-arch}.  
We develop a novel Rule Discovery Protocol (\Autoref{sec:rule-discov-prot}), 
which uses a method based on Monte-Carlo Tree search (MCTS) to search a rule grammar in a structured way and find population rules contained by the clients. 
The protocol follows the traditional MCTS steps (Selection, Expansion, Querying, and Backpropagation). To provide LDP, we adapt the querying phase to use randomized response.  
To find clinically useful rules, we adapt the MCTS scoring function, which guides the search process about which subtrees to continue searching down, to use privacy-preserving estimates of the number of clients who have rules that match a template rule structure in the grammar. 
By guiding the searching based on client responses, and taking advantage of the rule grammar, we are able to efficiently learn population rulesets including rules with complex structures.

Each query in the Rule Discovery Protocol is allocated a privacy loss budget that determines the randomized response noise used in the response. We develop an adaptive budget allocation method, which dynamically provisions the privacy  budget (\Autoref{sec:adaptive-budget-alloc}). The  intuition 
is to find the min. budget per query to gain enough information to determine if a node should be further explored.

We evaluate our protocol on three clinical datasets from different medical domains: 
Intensive Care Unit, 
Sepsis, 
and Diabetes, 
and find that we are able to learn population rulesets with high coverage and clinical utility, even at low privacy loss budgets (\Autoref{sec:exps}). 


%% file: 2.prelim.tex
\section{Preliminaries}\label{sec:prelim}

In this section we provide an overview of relevant background on Signal Temporal Logic, STL Learning (\Autoref{sec-prelim-stl}), Monte-Carlo Tree Search (\Autoref{sec-prelim-mcts}), and Local Differential Privacy  (\Autoref{sec-prelim-ldp}). 

\subsection{Signal Temporal Logic}\label{sec-prelim-stl}
\begin{figure}
\centering
\includegraphics[width=\linewidth]{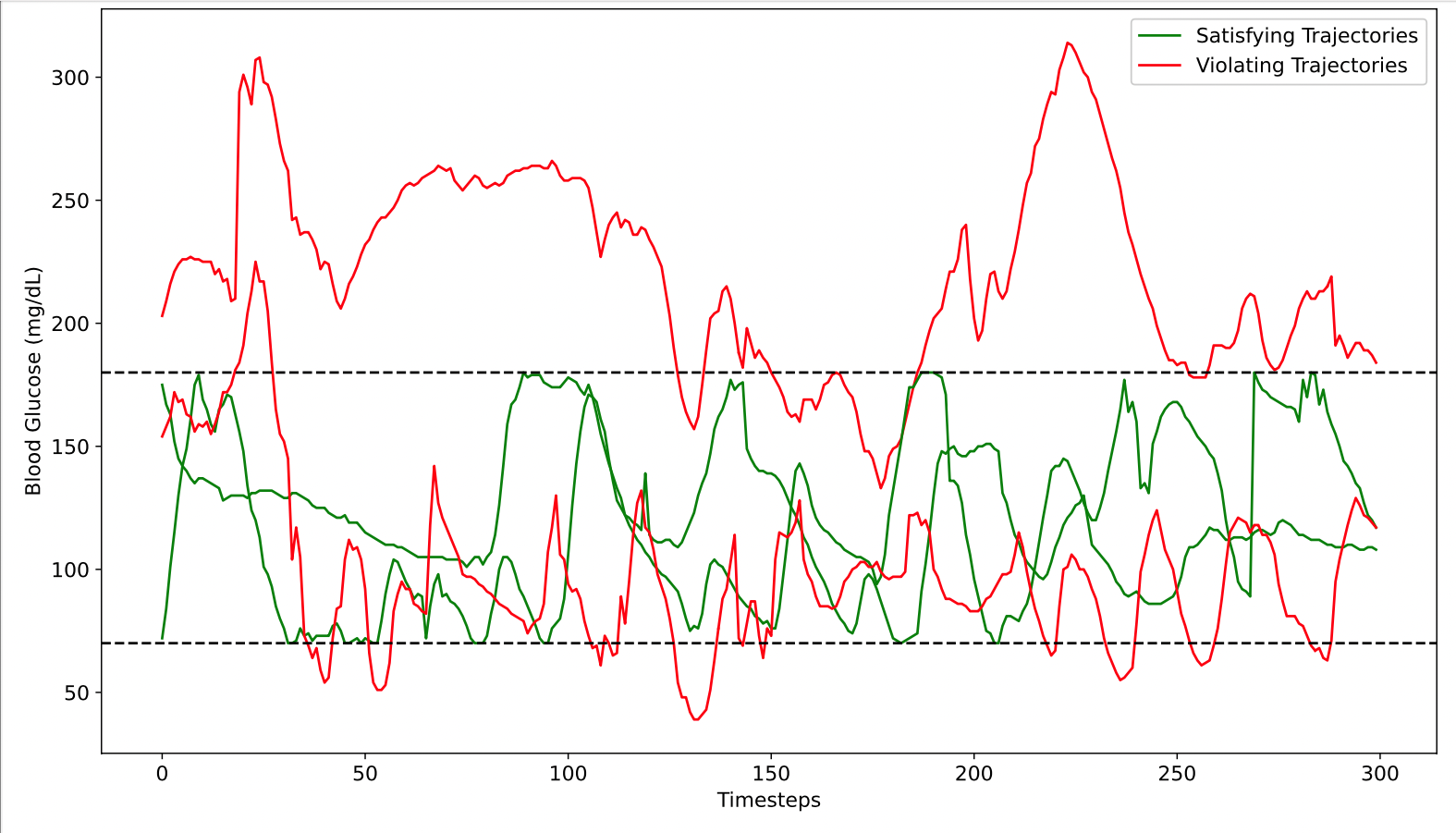}
\caption{Visual of the STL-learned rule \ruletemplate{$\square_{[0,300]}$(BG $\geq$ 70 $\land$ BG $\leq$ 180)} from glucose trajectories. The green trajectories satisfy the rule (glucose in range), and the red violate it.} 
\Description{}
\label{fig-t1d-rule}
\end{figure}

Signal Temporal Logic (STL) is a formal specification language used to express temporal properties over real-valued trajectories, commonly used to reason about behaviors of real-world systems, such as cyber-physical systems~\cite{bartocci2018specification}. 
We denote $Z$ and $P$ as finite sets of real and propositional variables. We let $w$ : $\mathbb{T} \longrightarrow \mathbb{R}^m \times \mathbb{B}^u$ be a multi-dimensional signal, where $\mathbb{T} = [0,d) \subseteq \mathbb{R}$, $m = |Z|$ and $u = |P|$. 
The syntax of an STL formula $\varphi$ over $Z \cup P$ is defined by the grammar:
$$\varphi ::= p \mid z \sim l \mid \neg \varphi \mid \varphi_1 \land \varphi_2 \mid \square_I \varphi \mid \Diamond_I \varphi \mid \varphi_1 \until_I \varphi_2$$
where $p \in P$, $z \in Z$, $\sim{} \in \{<, \leq\}$, $l \in \mathbb{Q}$, $I \subseteq \mathbb{R}^+$ is an interval and $\square$, $\Diamond$, and $\until$ denote temporal operators ``always'', ``eventually'', and ``until''. 
STL can be interpreted over a signal to describe the satisfaction of a formula (an example is in \Autoref{fig-t1d-rule}). Bartocci et al.\ \cite{bartocci2018specification} provide a comprehensive survey on STL and its use in cyber-physical systems. 

\shortsection{STL Learning} Although there are many rule-based machine learning methods, we focus on STL Learning due to its ability to expressively represent temporal properties of real-valued signal trajectories and because it is used in real clinical use cases (e.g.,~\cite{lamp2019logic}). 
STL learning takes advantage of the expressivity of the STL language and provides ML techniques to infer STL formulae and parameters from continuous  trajectories~\cite{bartocci2022survey}. There are many STL Learning algorithms, and our Rule Discovery Protocol does not depend on how the local client STL rules are learned. For our experiments, we use the Nenzi et al.~\cite{nenzi2018robust} genetic algorithm-based methodology as it is well suited to supervised classification tasks and is able to learn \emph{both} the parameters and the structure of STL formulae from real data. 
The algorithm requires positive and negative trajectories (e.g., regular and anomalous) and its goal is to learn rules that best characterize and separate 
the positive and negative samples. 
We use our own implementation of such an algorithm, developed in Python3 and available here: 
\url{https://github.com/jozieLamp/STLlearning}.
Some examples of learned STL rules are shown in \Autoref{table:stl-rules}.

\begin{table*}[t]
\caption{Example STL Rules Learned from Our Datasets}\label{table:stl-rules}
\vspace{2pt}
\centering
\resizebox{0.75\linewidth}{!}{%
\begin{tabular}{c|l} \hline \toprule
\textbf{Dataset} & \textbf{Rule} \\ \midrule
\multirow{3}{*}{ICU} & \ruletemplate{$\square_{[0,0]}$(HR $\geq$ 80.369 $\land$ Pulse $\geq$ 74.034)}  \\ \cmidrule{2-2}
 & \ruletemplate{((MET $\geq$ 0.007) $\until_{[0,1]}$ (DeathProb = 0.032))} \\ \cmidrule{2-2}
 & \ruletemplate{$\Diamond_{[0,1]}$(Blood\_Urea\_Nitrogen $\leq$ 12.889 $\land$ Creatinine $\geq$ 0.723)}  \\ \midrule
\multirow{3}{*}{Sepsis} &  \ruletemplate{((Temp $\geq$ 9.059) $\until_{[1,1]}$ (BaseExcess $\geq$ 0.048))} \\ \cmidrule{2-2}
 & \ruletemplate{$\square_{[0,1]}$(((HGB $\geq$ 0.385 $\land$ MAP $\leq$ 110.015) $\lor$ Bilirubin\_Direct $\leq$ 107.835) $\land$ AST $\geq$ 47.955)}  \\ \cmidrule{2-2}
 & \ruletemplate{$\Diamond_{[1,2]}$(((PaCO\_2 $\geq$ 0.171 $\lor$ Chloride $\leq$ 8.029) $\implies$ Potassium $\geq$ 0.014) $\land$ SepsisProb $\geq$ 0.85)} \\ \midrule
 \multirow{3}{*}{T1D} & \ruletemplate{((HbA1c $\geq$ 7.571) $\until_{[1,2]}$ (Hypoglycemia = 1.0))} \\ \cmidrule{2-2}
 & \ruletemplate{$\square_{[0,1]}$((TotalDailyInsPerKg $\leq$ 0.305 $\implies$ PtA1cGoal $\geq$ 0.518\%) $\land$ GFR $\leq$ 89\%)} \\ \cmidrule{2-2}
 & \ruletemplate{$\Diamond_{[1,1]}$(((BMI $\geq$ 27.066 $\lor$ HeightCm $\geq$ 180.022) $\implies$ HbA1c $\leq$ 6\%) $\land$ Bolus $\geq$ 57.424)} \\  \bottomrule
\end{tabular}
}
\captionsetup{width=0.75\linewidth}
\caption*{\scriptsize The ICU rules characterize relationships between labs, physiological values and mortality (MET, DeathProb). The Sepsis rules characterize relationships between lab values 
and sepsis outcomes (SepsisProb). The T1D rules characterize relationships between insulin, blood glucose levels, glomular filtration rate (GFR), body mass index (BMI) and glycemic outcomes (HbA1c levels, goals and hypoglcemia).}
\end{table*}

\subsection{Monte-Carlo Tree Search (MCTS)}\label{sec-prelim-mcts}
Monte-Carlo Tree Search (MCTS)~\cite{browne2012survey, kocsis2006bandit} is a well-known algorithmic search method used to solve sequential decision problems and search large combinatorial spaces. 
MCTS 
works by building a search tree that balances \emph{exploration}, finding new options in the search space, and \emph{exploitation}, focusing on the parts of the space that are most likely to return good rewards. There are four key phases of MCTS: Selection, Expansion, Querying (traditionally called Simulation), and Backpropagation.

A common MCTS algorithm is the Upper Confidence Bounds for Trees (UCT) algorithm~\cite{kocsis2006bandit}. 
This method 
asymmetrically searches the tree, focusing on the pathways that are most promising. The UCT scoring function, which we adapt for our method, is:
$$\mathit{score} = \mathit{rw} + C_p \times \sqrt{\frac{v_{\mathit{parent}}}{v_{b}}}$$
where $\mathit{rw}$ is the current reward, $v_{\mathit{parent}}$ is the visit count of the parent node, and $v_b$ 
is the visit count of the current node. The hyperparameter
$C_p$ balances the exploration vs.\ exploitation trade-off in the MCTS search, and is typically set to $\frac{1}{\sqrt{2}}$. 

\subsection{Local Differential Privacy}\label{sec-prelim-ldp}
Local Differential Privacy (LDP) is a paradigm well suited to the distributed framework deployed for many CDSS systems. It provides privacy assurances to clients without relying on any external server since individual users each perturb their own data before it is collected and aggregated~\cite{kasiviswanathan2011can}. 
In this setting, a centralized, untrusted server $S$, wishes to aggregate some summary statistic $s$ from $n$ individual clients' data records, $\{x_1, ...,x_n\}$, that contain private information. 
Each client locally perturbs the requested data $x_i$ 
before sending it to the server. 
An algorithm $\mathcal{A}$ satisfies $\epsilon$-local differential privacy where $\epsilon>0$ if, for any possible pairs of inputs $x$ and $x'$:
$$\forall s \in \mathit{Range}(\mathcal{A}): \frac{Pr[\mathcal{A}(x) = s]}{Pr[\mathcal{A}(x') = s]} \leq e^\epsilon$$
where $\mathit{Range}(\mathcal{A})$ denotes every possible output of $\mathcal{A}$. 

%% file: 3.approach.tex
\begin{figure*}[t]
\centering
\includegraphics[width=0.75\linewidth]{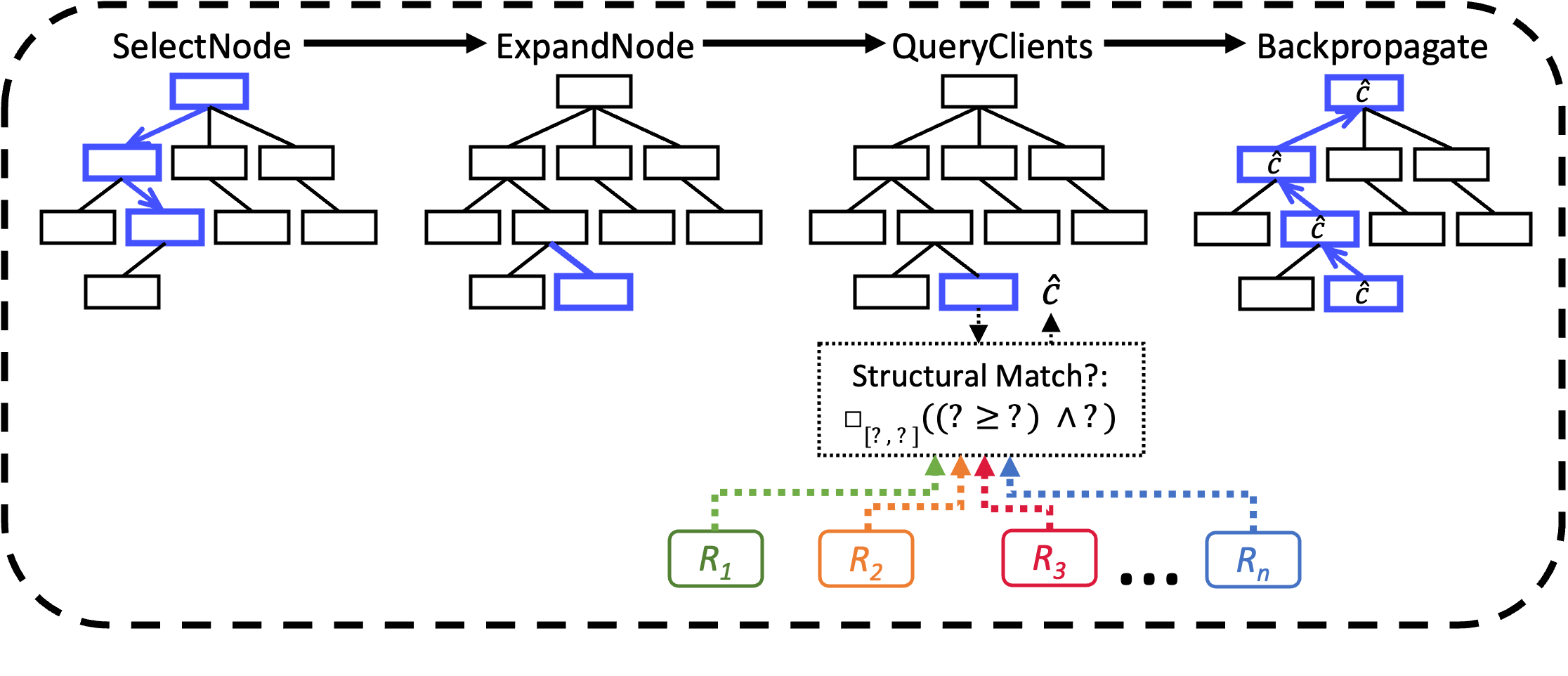}
\caption{Rule Discovery Protocol. The protocol iterates through each MCTS phase (SelectNode, ExpandNode, QueryClients, Backpropagate) to send a series of structured queries to the clients, who respond using randomized response, to generate 
$R_S$.} 
\Description{}
\label{fig-protocol}
\end{figure*}

\begin{figure*}[t]
\centering
\includegraphics[width=\linewidth]{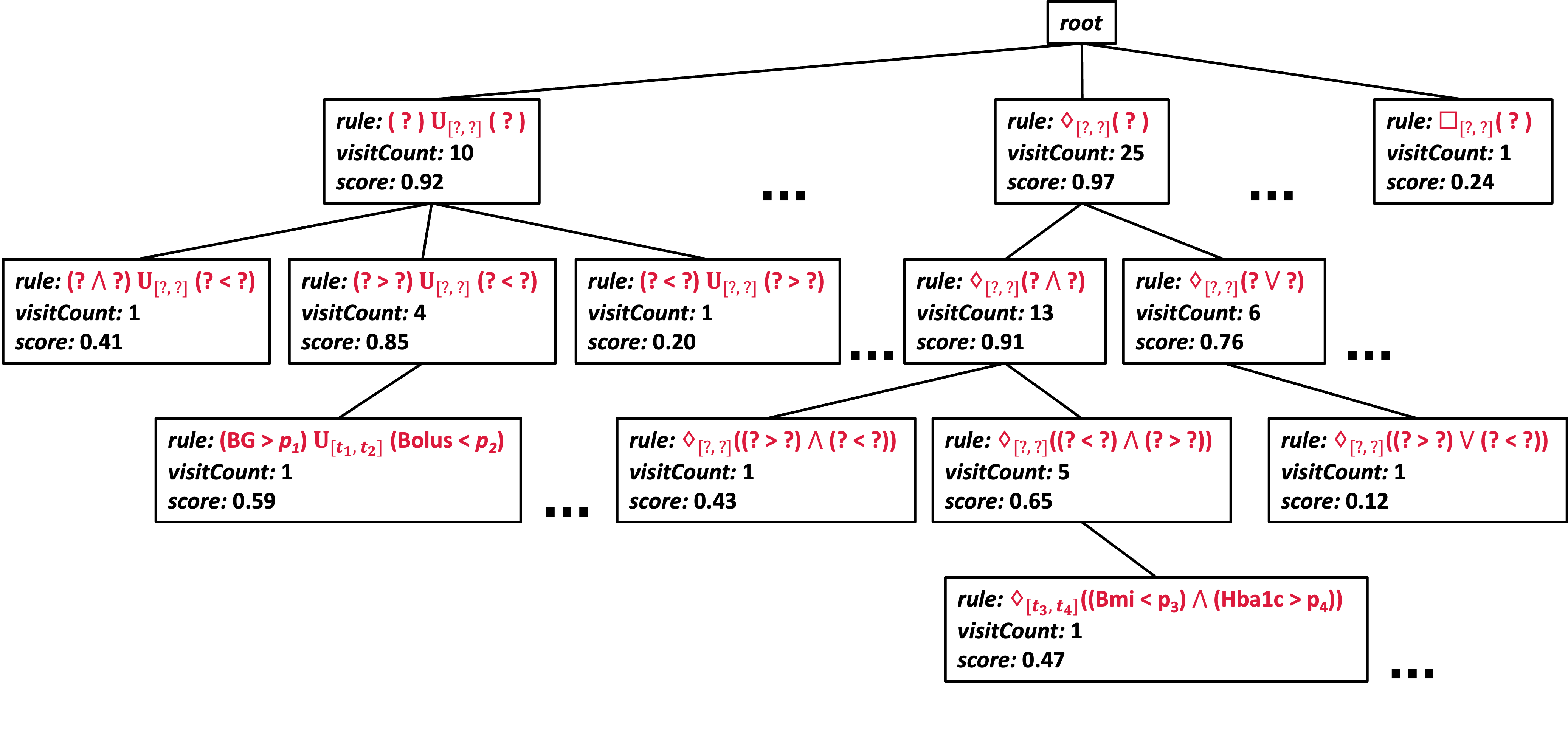}
\caption{Example Partial Exploration Tree. Tree nodes contain the rule and MCTS components \textit{visitCount} and \textit{score}. 
} 
\Description{}
\label{fig-exploration-tree}
\end{figure*}

\section{Rule Discovery Protocol}\label{sec:approach}
We introduce a rule discovery protocol (\Autoref{fig-protocol}) which integrates MCTS with LDP to search a rule grammar and find population rules.  
We walk through each protocol phase in \Autoref{sec:rule-discov-prot}--\Autoref{sec:backprop}. 
Then, \Autoref{sec:adaptive-budget-alloc} describes an adaptive privacy loss budget allocation method to determine the privacy loss budget for each query.

\subsection{Overview}\label{sec:rule-discov-prot}

The rule discovery protocol uses an \textit{exploration tree}, $T$, to search over an STL grammar $G$. The protocol follows the traditional MCTS steps (Selection, Expansion, Querying, 
and Backpropagation), adapted to support LDP by using randomized response when clients respond to queries. 
We use an MCTS-based approach for our protocol due to its ability to efficiently search a tree structure while balancing the trade-off between exploration (i.e., finding new nodes and pathways through the tree), and exploitation (i.e., focusing on the known nodes in the tree that maximize the score function). This design is advantageous in an LDP setting where the number and accuracy of the queries is limited by the privacy budget. 

\shortsection{Exploration Tree} An example partial exploration tree is shown in \Autoref{fig-exploration-tree}. Each node records MCTS properties including node visit count and current score, which indicates its priority for exploration, explained in \Autoref{ssec:selectnode}, as well as the rule structure. 
In a completed exploration tree, internal nodes contain incomplete rule templates where unfinished parts of the rules are represented using "{\sf ?}". Leaves contain either incomplete rule templates (indicating that a tree path was not fully explored and that node's children have not been visited), or completed rules. The completed rules constitute the learned population ruleset $R_S$. The rule discovery protocol searches $G$ to iteratively build the exploration tree $T$ and return the population rule set $R_S$. 

\shortsection{Threat Model}
We assume the network traffic is not observable to an adversary; our focus is on providing client privacy from the central server. In a setting where network traffic is exposed, it would be necessary to modify the protocol to ensure the communication pattern, timing, and packet sizes do not leak information about a client's rules. For our differential privacy notion, we quantify the unit of privacy as one rule, and we assume all rule structures are independent. In practice, a client may learn multiple rules that convey the same privacy information, so a privacy guarantee at the level of individual rules as the unit of privacy would be insufficient. 

We assume all participating clients are honest---they follow the protocol as prescribed, keeping track of their own privacy loss budget, implementing randomized response as intended, and refusing to respond to any more queries once their privacy loss budget is expended. To keep things simple in our design, 
we assume all clients have the same privacy loss budget and each query uses the same per-query budget for all clients. An adversarial client could respond to queries in ways that would compromise the results, but we assume that in relevant clinical settings all participants would be motivated for the aggregate model to be as useful as possible.

\begin{algorithm}[t]
\caption{Rule Discovery Protocol}
\SetAlgoLined
\DontPrintSemicolon
\SetNoFillComment
\SetKwProg{Fn}{protocol}{:}{end}
\SetKwProg{SubFn}{Sub-Function}{:}{end}
\SetKwFunction{FMain}{DiscoverRules}
\SetKwFunction{selection}{SelectNode}
\SetKwFunction{expansion}{ExpandNode}
\SetKwFunction{querying}{Query}
\SetKwFunction{backprop}{Backpropagate}
\SetKwFunction{pruneTree}{PruneTree}
\SetKwFunction{selectchildnode}{selectChildnode}
\SetKwFunction{adptvProt}{Adaptive Budget Protocol}
\SetKwFunction{findCutoff}{Prune\_Branch}
\SetKwFunction{CDF}{CDF}
\SetKwFunction{findBeta}{allocatePrivacyBudget}
\SetKwFunction{errorProb}{Compute\_Error\_Prob}
\SetKwFunction{query}{SendQuery}
\SetKwFunction{returnLeaves}{returnVisitedLeaves}
\label{alg-protocol-overview}
\Fn{\FMain{$G$ {\rm (rule grammar)}, 
$\mathcal{V}$~{\rm (valid rule threshold)}, $\epsilon$~{\rm (privacy loss budget)}, $n$~{\rm (number of clients)}, $\theta$~{\rm (exploration threshold)}}}{
    $R_S \longleftarrow \emptyset$\;
    $T \longleftarrow$ \emph{empty exploration tree}\;
    $b \longleftarrow$ $T.root$\; 
    $\mathit{plb} \longleftarrow$ $\epsilon$
   
    \While{$\mathit{plb}$ > 0} {
        $b_{selected}$ $\longleftarrow$ \selection{b, $T$, $G$}\; \label{algline-select}
        $b$ $\longleftarrow$ \expansion{$b_{selected}$, $T$, $G$}\; \label{algline-expandbegin}
        $\hat{c}$, $\mathit{plb}$ $\longleftarrow$ \querying{$b$, $\mathit{plb}$, $\epsilon$, $\mathcal{V}$, $n$, $\theta$, $R_S$}\; \label{algline-query}
        \backprop{$b$, $\hat{c}$, $T$}\; \label{algline-backprop}

    }
    \Return population ruleset $R_S$\; \label{algline-return}
}
\end{algorithm}

\shortsection{Protocol Algorithm} 
The protocol, described in \Autoref{alg-protocol-overview}, takes as input a rule grammar $G$; the valid rule threshold $\mathcal{V}$, the fraction of clients who must have a match to the rule structure for it to be considered viable in the protocol; the privacy loss budget $\epsilon$; 
the number of clients $n$; and an exploration threshold $\theta$. Details about $\theta$ are explained in \Autoref{sec:adaptive-budget-alloc}. 
In the start of the algorithm, the population ruleset $R_S$ and exploration tree $T$ are initialized, the current node $b$ is set to the root of the tree, and a variable tracking the amount of privacy budget used, $plb$ is initialized to the global privacy loss budget $\epsilon$.
The protocol loops iteratively through the four MCTS phases until the privacy loss budget has been used and the aggregated population ruleset $R_S$ is returned. 

To simplify protocol design, we assume all clients have the same privacy budget, and every query is sent to every client. We also assume query executions are done completely; there are additional opportunities to save privacy loss budget by cutting off a query once enough responses have been received.  If these simplifying constraints were removed, there are many opportunities to use the privacy loss budget more efficiently, such as querying subsets of clients or adjusting the privacy loss budget for a query as more info is learned from clients.
We describe each protocol phase next.



\subsection{Selection}\label{ssec:selectnode}
In the first MCTS phase, the protocol selects a node to explore (Alg.~\ref{alg-protocol-overview}, \Autoref{algline-select}). This function follows the traditional Selection implementation in MCTS, and pseudocode is available in \Autoref{apdx:addtl-algs}. The next node is recursively selected by either choosing the child node of the current node $b$ that is unvisited, or choosing the child node that returns the maximum score according to the scoring function, discussed next. 
The selection function returns when a terminal node is reached.

\shortsection{Scoring} 
Scoring uses the classical UCT score~\cite{kocsis2006bandit}, with the reward adapted to be the percent of clients who have a match to the rule structure (received from the clients' randomized responses and explained below in QueryClients).
For node $b$,
\begin{equation}\label{eq-slct-scoring}
    score =\begin{cases}
    0, \qquad & \textit{If } \frac{\hat{c}}{n} < \mathcal{V}\\
    \frac{\hat{c}}{n} + C_p \times \sqrt{\frac{v_{parent}}{v_{b}}}, \qquad & \textit{otherwise}
    \end{cases}
\end{equation}
where $\hat{c}$ is the client match count, $C_p$ is a hyperparameter balancing exploration vs. exploitation in MCTS, $v_{parent}$ is the visit count of the parent node, and $v_b$ is the visit count of the current node.

\subsection{Expansion}
Next, in the second phase the protocol expands reachable nodes and adds them to the exploration tree (Alg.~\ref{alg-protocol-overview}, \Autoref{algline-expandbegin}). This function follows the traditional MCTS Expansion implementation in MCTS, and pseudocode is available in \Autoref{apdx:addtl-algs}. It either just returns $b_{\mathit{selected}}$, or chooses from among the child nodes according to a \textit{selection policy}. This policy may select a node to expand \textit{randomly} or based on the node that has the highest score (\Autoref{eq-slct-scoring}). 

\subsection{Querying}
In the next phase, clients are queried using randomized response (Alg.\ \ref{alg-protocol-overview}, \Autoref{algline-query}).
This step is classically known as Simulation; our adaptation is illustrated in \Autoref{alg-querying}.

\shortsection{Allocate Privacy Budget}
The local privacy loss budget, $\beta$, to be used by each client for the query, is determined (Alg.\ \ref{alg-querying}, \Autoref{algline-allocbudget}). 
In the baseline method, a uniform budget is used for every query by just dividing the total budget by a pre-specified number of queries: 
$\beta = \epsilon/Q$, where $Q$ is the number of queries. 
For the adaptive method, $\beta$ is determined using the method described in \Autoref{sec:adaptive-budget-alloc}. 

\begin{algorithm}[t]
\caption{Query Clients for Matching Rules}
\label{alg-querying}
\SetAlgoLined
\DontPrintSemicolon
\SetNoFillComment
\SetKwProg{Fn}{protocol}{:}{end}
\SetKwFunction{getQuery}{Query}
\SetKwFunction{querymatch}{QueryRuleMatch}
\SetKwFunction{queryparam}{QueryParameters}
\SetKwFunction{parambudget}{ParamPrivacyBudget}
\Fn{\getQuery{$b$, $\mathit{plb}$, $\epsilon$, $\mathcal{V}$, $n$, $\theta$, $R_S$}}{
    $\beta \longleftarrow $ \findBeta{$\mathit{plb}$, $\epsilon$, $\mathcal{V}, n, \theta$} \; \label{algline-allocbudget} 

    $t \longleftarrow$ $b.rule$ \tcp{Get current rule structure} \label{algline-getrule}
    \For{$cl$ {\bf{in}} clients}{ \label{algline-qstart}
            \tcp{Query each client for structural rule match to $t$}
            $y$ += $cl$.\querymatch{$t, \beta$} 
        } \label{algline-qend}
    $\mathit{plb}$ $\longleftarrow$ $\mathit{plb} - \beta$ \tcp{Update used budget} \label{algline-addplb}
    $\hat{c} \longleftarrow \frac{y - nq}{p-q}$ \tcp{Unbiased estimate of count} \label{algline-unbscnt}

    \If{$t$ {\rm is complete}}{ \label{algline-paramq-start}
        \tcp{Determine privacy budget for learning parameters}
        $\beta_{\mathit param} \longleftarrow$ \parambudget($t$, $\mathit{plb}$) \label{algline-parambudg}

        $\mathit{plb}$ $\longleftarrow$ $\mathit{plb} - \beta_{param}$ \tcp{Update used budget}\label{algline-paramupdbudg}
        $b.rule$ $\longleftarrow$ \queryparam{$b.rule$,$\beta_{\mathit{param}}$}\; \label{algline-queryparams}
        $R_S.insert(b.rule)$ \tcp{Add completed rule to $R_S$}
    }\label{algline-paramq-end}

    \Return{$\hat{c}$, $\mathit{plb}$}\label{algline-qreturn}
}
\end{algorithm}

\shortsection{Querying}
$\beta$ is used to send a query (in the form of a rule template) to all the clients and obtain an estimate of how many clients have a match to the rule structure contained at the selected node. 
The function first gets the partial rule template $t$ from the current node $b$ (Alg.\ \ref{alg-querying}, \Autoref{algline-getrule}). 
Next, it queries each client to get the number of yes responses, $y$, who have a match to $t$ (\Autoref{algline-qstart}--\Autoref{algline-qend}). Each client gives their binary (yes/no) response following a Direct Encoding Randomized Response method~\cite{wang2017locally}. Additional details about the rule matching process are explained below. Then, the used privacy loss budget is updated (\Autoref{algline-addplb}) and the unbiased estimate of the count $\hat{c}$ is computed following randomized response~\cite{wang2017locally} (\Autoref{algline-unbscnt}). 
$\hat{c}$ is used to inform the score function (\Autoref{eq-slct-scoring}) to guide the protocol in determining whether or not it should continue searching down a pathway. 
If a complete rule is found (one without any "{\sf ?}"), the parameters of the rule are queried (described below) and the rule is added to $R_S$ (\Autoref{algline-paramq-start} -- \Autoref{algline-paramq-end}). Finally, $\hat{c}$ and $\mathit{plb}$ are returned (\Autoref{algline-qreturn}).


\begin{figure}
\centering
\includegraphics[width=0.8\linewidth]{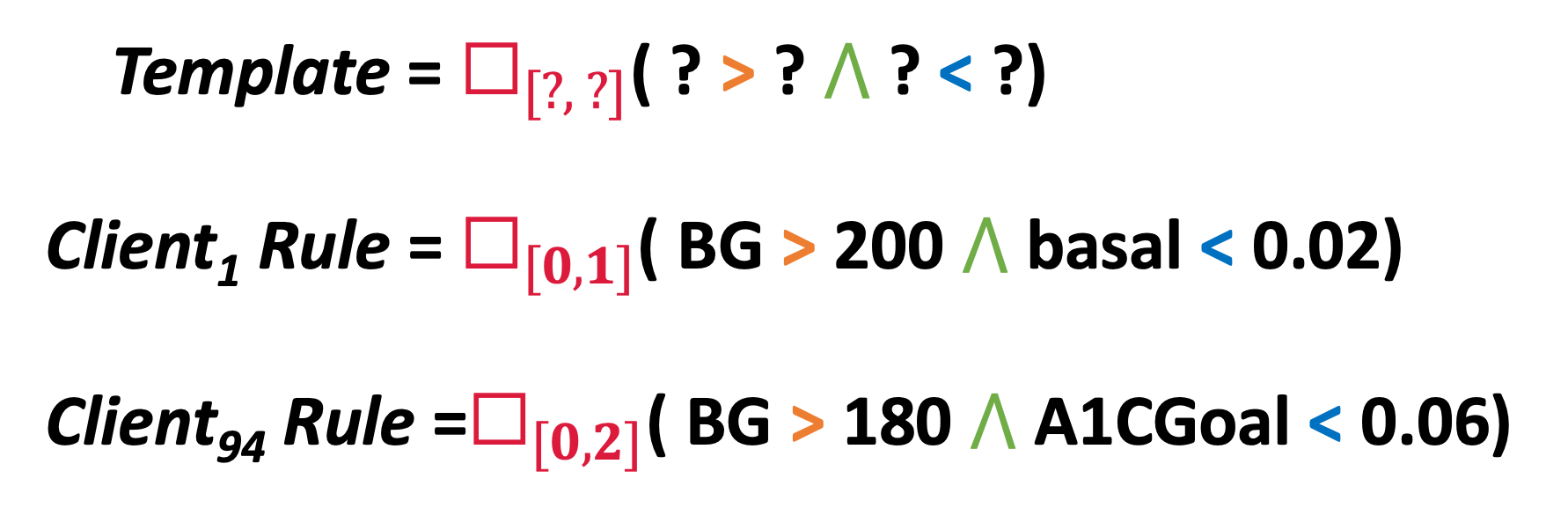}
\caption{Example Rule Matching. Colors indicate the part of the rule to be matched. In the template, the variables have not yet been specified (part of the {\sf ?}s), so the template matches client rules with different variables.} 
\Description{}
\label{fig-ex-rule-match}
\end{figure}

\shortsection{Client Rule Matching} 
In the query, the server sends out the rule template $t$ to all the clients. The clients each check to see if they have any rules that contain a syntactic match to the template. A syntactic match is a structural rule match, in which the specified parts of $t$ have matches to the client rule, and all other parts of the client rule (i.e., the "{\sf ?}" in $t$) are ignored. To account for possible semantic matches (equivalence relations following the defined STL logic~\cite{bartocci2018specification}, in which the client rule semantically has the same meaning as the template even though the syntactic structure of the rules may differ), we assume that all rule structures have been converted to a canonical set 
in the rule learning. 
\Autoref{fig-ex-rule-match} shows an example of two client rules matching a template.


\shortsection{Parameter Querying}
If a leaf node in the exploration tree has been reached (representing a completed rule structure, one without any "{\sf ?}" marks), the parameters of the discovered rule structure are queried and aggregated as well. We allocate the parameter budget $\beta_{param}$ by using a small fixed constant multiplied by the number of parameters there are in the rule structure to fill in (Alg.\ \ref{alg-querying}, \Autoref{algline-parambudg}). After updating the used budget $plb$ (\Autoref{algline-paramupdbudg}), the clients are queried for their parameters using $\beta_{param}$ (\Autoref{algline-queryparams}). If a client does not have a rule match to $t$, they respond with random (noised) parameter values. 
We aggregate parameters using a standard mean value estimation process using LaPlacian noise~\cite{wang2017locally}. To aggregate the parameters, a percentile threshold $\tau$ is given, and a parameter value is selected at or below which (inclusive) $\tau$\% of the scores in the distribution may be found.

\subsection{Backpropagation}\label{sec:backprop}
In the last MCTS phase, (Alg.~\ref{alg-protocol-overview}, \Autoref{algline-backprop}) scores are propagated up the exploration tree.
This follows traditional Backpropagation in MCTS (pseudocode is available in \Autoref{apdx:addtl-algs}). Starting at the current node and continuing up through the node's parents (until the root node is reached), each node updates the following: the match count $\hat{c}$ is added to $b.responses$, a list tracking the previous yes responses, the number of visits $b.visitCount$ is incremented and the score of the current node, $b.score$ is updated using the scoring method (\Autoref{eq-slct-scoring}). 


\subsection{Adaptive Budget Allocation}\label{sec:adaptive-budget-alloc}
We detail next how the privacy loss budget is dynamically allocated in the adaptive method. 
We define $c$ as the true (unknown) count of how many clients have a match to the rule structure ($t$) at the current node ($b$) and $\hat{c}$ as the estimated count (obtained from noised client responses in the protocol).  
Based on the score function, any nodes that have client match counts $\hat{c}$ below the valid rule threshold $\mathcal{V}$ are ignored (not explored) in the searching, since they are unlikely to have clients with rule matches.  
As a result of noise being added to the client responses, there are two types of error that can occur in the searching: (1) Wasting queries searching down pathways which few clients have matches to ($\frac{\hat{c}}{n} \geq \mathcal{V}$ but there is no valid rule in the subtree)
and (2) failing to explore parts of the grammar that contain valid rules ($\frac{\hat{c}}{n} < \mathcal{V}$ but the subtree contains a rule where $\frac{c}{n} \geq \mathcal{V}$).  
We prioritize avoiding the second type of error, as sending a few more queries than necessary is better than missing entire subtrees of the grammar that may contain important and relevant rules.

\begin{figure*}[t]
     \centering
     \subfigure[ICU]{
         \centering
         \includegraphics[width=0.325\textwidth]{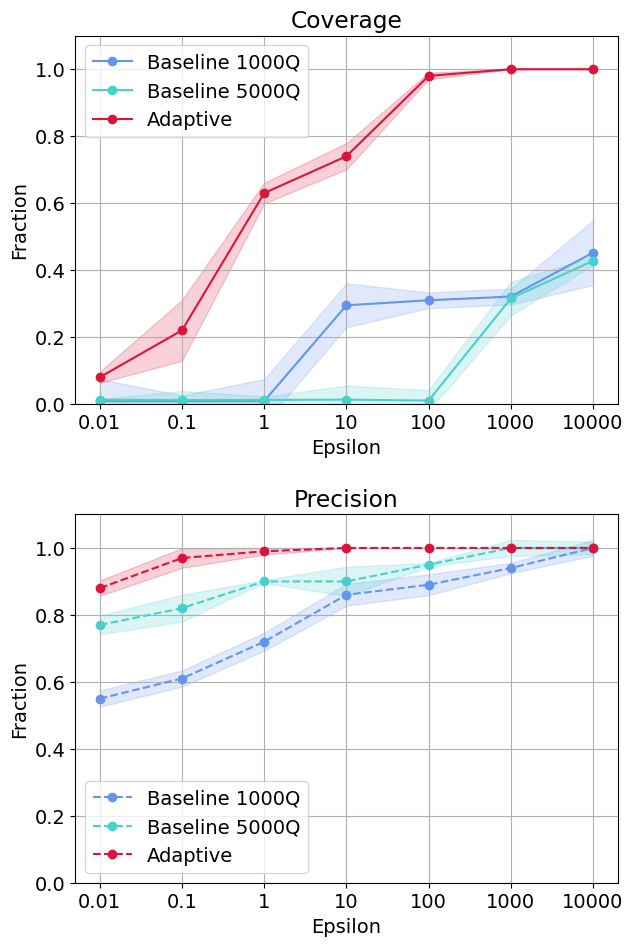}}
     \hfill
     \subfigure[Sepsis]{
         \centering
         \includegraphics[width=0.325\textwidth]{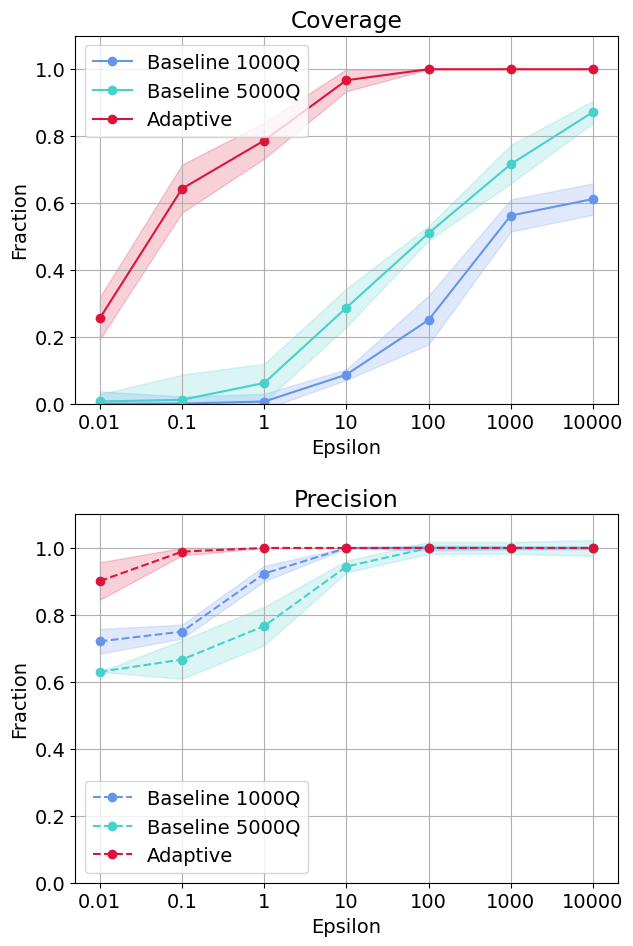}}
    \hfill
    \subfigure[T1D]{
         \centering
         \includegraphics[width=0.325\textwidth]{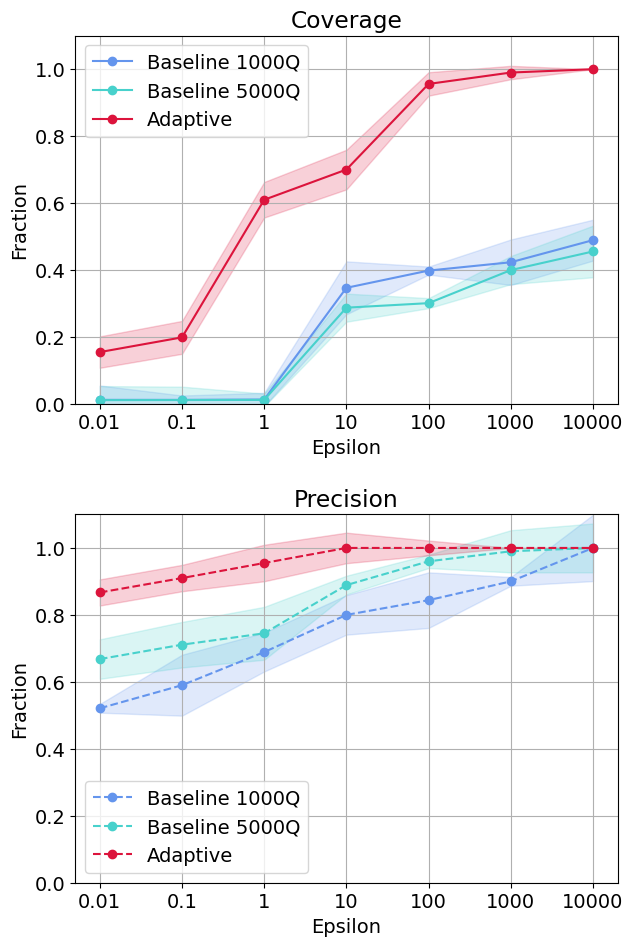}}
    \caption{Coverage ($\mathcal{V}=1\%$, $\theta=5\%$)}
    \label{fig:cov-vs-base}
    \Description{}
\end{figure*}

To this end, we adaptively allocate our budget by finding the minimum budget per query, $\beta$, that ensures the probability of failing to explore a subtree that is likely to have valid rules is bounded by a user-specified exploration trade-off threshold, $\theta$. 
Following typical LDP randomized response~\cite{wang2017locally}, a user outputs a response equal to the true response 
with probability $p$ and the negation of the true response 
with probability $q$:
\begin{align}
    &p = \frac{e^\beta}{1 + e^\beta} \\
    &q = 1-p = \frac{1}{1 + e^\beta}
\end{align}
Given $p$ and $q$, we can compute the estimated match count to a query, $\hat{c}$, as:
\begin{equation}
    \hat{c} = y \times p + (n - y) \times q 
\end{equation}
where $y$ is the number of yes responses returned from the random response mechanism. 
To find $\beta$, we formulate an optimization problem as follows:
\begin{align}
    &\min_{\beta} \left( \int_{y=0}^{n} (P(\frac{\hat{c}}{n} < \mathcal{V} \;\lvert\; \frac{c}{n} = \mathcal{V}) ) \right) \leq \theta 
\end{align} 
Since we have not actually sent a query yet, we do not have any responses from the clients and do not know the value of $y$. Therefore, we iterate over all possible values of $y$ from 0 to the population size $n$. 
We assume the worst case scenario where the true percent $\frac{c}{n}$ is directly at the valid rule threshold $\mathcal{V}$. 
In summary, this equation seeks to find the minimum $\beta$, where, for all possible values of $y$, the probability that we falsely ignore this branch of the grammar is bounded by 
$\theta$.

%% file: 4.experiments.tex
\begin{table*}[t]
\caption{Clinical Dataset Details}\label{table-data-overview}
\centering
\resizebox{\linewidth}{!}{%
\begin{tabular}{ccccccccc} \hline
\toprule
\multicolumn{7}{c}{} & \multicolumn{2}{c}{\textbf{Negative Outcome}} \\ 
\textbf{Dataset} & \textbf{\# Patients} & \textbf{\# Features} & \textbf{Temporal Recording} & \textbf{\# Timepoints} & \textbf{Ave. \# Timepoints/Patient} & \textbf{Label} & \textbf{\% Patients} & \textbf{\% Timepoints} \\ \midrule
ICU & 8000 & 57 & Every 15 minutes & 2,437,318 & 304.70 & Deterioration & 1.59 & \num{5.21e-05}\\ 
Sepsis & 40,336 & 35 & Hourly & 1,552,210 & 38.48 & Sepsis & 1.06 & \num{2.74e-04} \\ 
T1D & 34,013 & 40 & Yearly & 140,461 & 4.13 & Hypoglycemia & 5.26 & 1.27 \\ \bottomrule
\end{tabular}
}
\end{table*}

\begin{table}[t]
\caption{STL-Learned Ruleset Characteristics}\label{table-rules-overview}
\centering
\resizebox{\linewidth}{!}{%
\begin{tabular}{cccccccc} \toprule
\textbf{Dataset} & \textbf{\# Client Rules} & \textbf{Ruleset Size} & \textbf{Rules/Patient} &  \textbf{Operators/Rule} \\ \midrule
ICU & 598,699 & 34,208 & 74.85 & 2.51 \\ 
Sepsis & 4,420,910 & 2,344,179 & 109.60 & 2.97\\ 
T1D & 2,105,755 & 1,353,598 & 61.91 & 4.35 \\ \bottomrule
\end{tabular}
}
\end{table}

\begin{figure*}[t]
     \centering
     \subfigure[ICU]{
         \centering
         \includegraphics[width=0.325\textwidth]{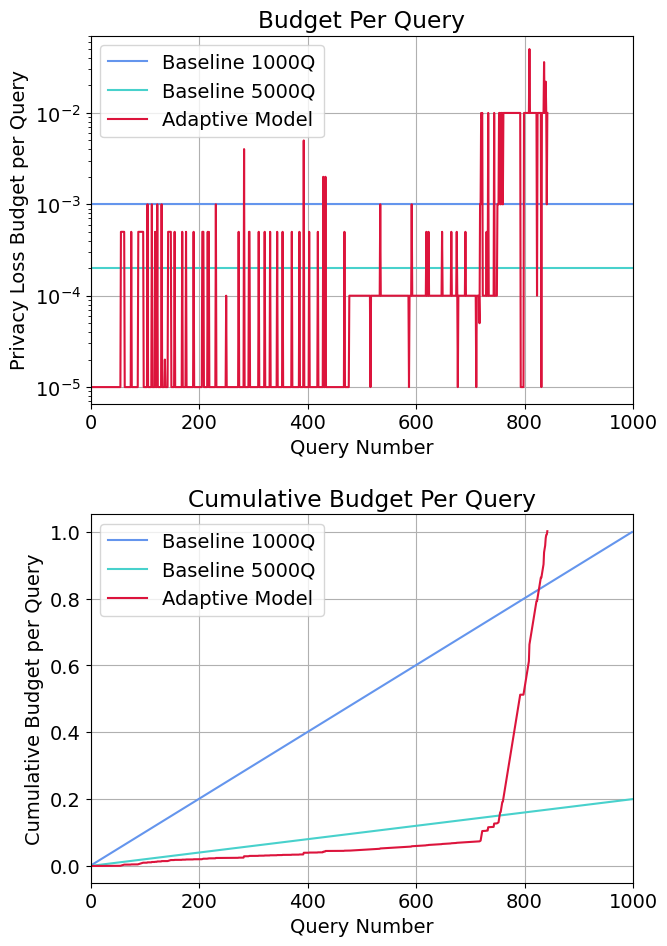}}
     \hfill
     \subfigure[Sepsis]{
         \centering
         \includegraphics[width=0.325\textwidth]{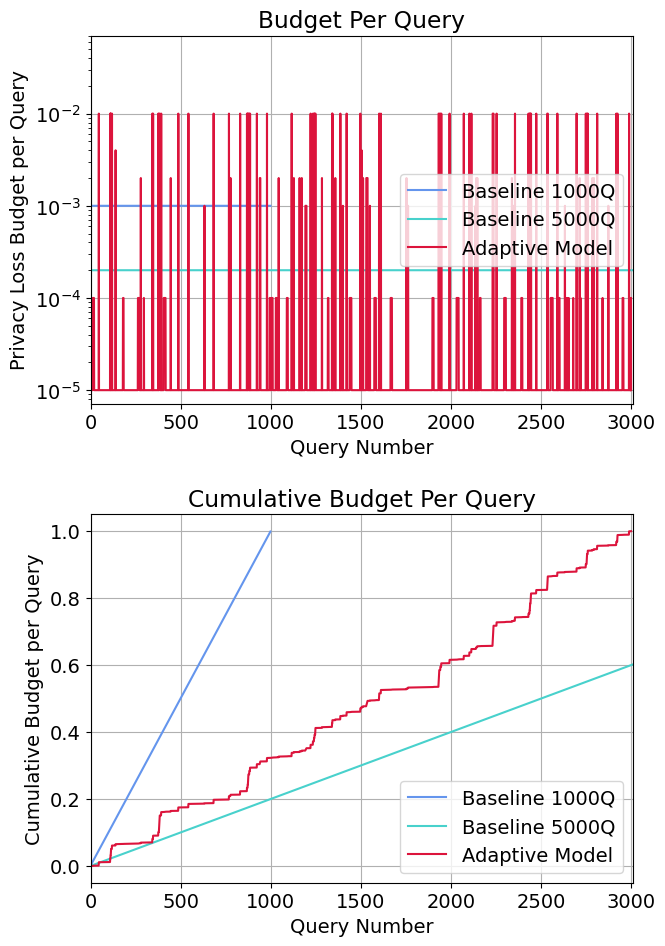}}
    \hfill
    \subfigure[T1D]{
         \centering
         \includegraphics[width=0.325\textwidth]{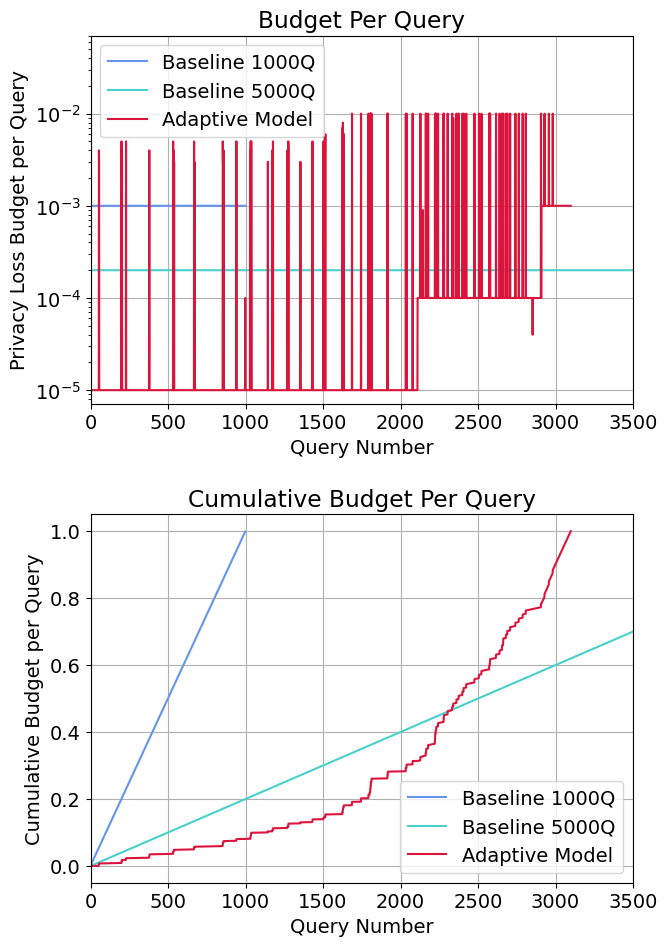}}
    \caption{Query Analysis ($\epsilon=1$, $\mathcal{V}=1\%$, $\theta=5\%$). The x-axis is truncated at the max \# of queries for the adaptive protocol in each graph to zoom in on interesting adaptive phenomenon (but the Baseline 5000Q lines continue uniformly to 5000 queries).}
    \label{fig:cov-query-analysis}
    \Description{}
\end{figure*}

\section{Experimental Evaluation}\label{sec:exps}

This section reports on the empirical evaluation of our framework. We first introduce the experimental setup (\Autoref{sec:exp-rulesets}), and then evaluate our framework based on two criteria: coverage (\Autoref{sec:exp-cov}) and clinical utility (\Autoref{sec:exp-util}). 


\subsection{Experimental Setup}\label{sec:exp-rulesets}
\shortsection{Data}
Three different open source datasets were chosen to evaluate the applicability of our approach for different clinical use cases (e.g., across different domains and patient populations).
Open data is necessary for reproducibility and means there are no actual privacy concerns with these data, but they are still representative of many sensitive and private datasets in the clinical setting.

An overview of the datasets' characteristics are shown in \Autoref{table-data-overview}. 
The Intensive Care Unit (ICU) dataset is from a study by Moss et al.~\cite{moss2017cardiorespiratory} predicting inpatient deterioration. 
The Sepsis dataset~\cite{reyna2019early} is from the
PhysioNet/Computing in Cardiology Challenge 2019, in which they were trying to develop better algorithms for early detection of sepsis using physiological trace data. 
The Type I Diabetes dataset (T1D)~\cite{beck2012t1d}, comes from the T1D Exchange Registry and collects longitudinal information of patients with T1D at each routine annual clinic exam between July 2007 to April 2018. 
We used the ICU dataset in developing our methods and for both validation and testing purposes, but only used the Sepsis and T1D datasets for final testing and evaluation to simulate a realistic scenario where the method cannot be tuned to particular data but must be determined without access to the intended data. 

\shortsection{Rules} 
A set of rules was learned locally for each patient in each dataset using STL Learning. The STL Learners were trained for 1000 epochs to predict each dataset's outcome (Deterioriation, Sepsis, and Hypoglycemia for the ICU, Sepsis and T1D datasets respectively). 
No limits were set on the number of rules outputted from each learner, so clients have different numbers of rules in their local rulesets.
Example learned rules are in \Autoref{table:stl-rules}. \Autoref{table-rules-overview} reports ruleset characteristics; \# Client Rules indicates the sum of the lengths of all individual client rulesets and Ruleset Size indicates the length of the total set of client rules (which has no duplicate client rules). 
\Autoref{fig:rule-breakdown} in \Autoref{appdx:eval} shows the breakdown of how many rules there are at different population percentages for each ruleset.

\shortsection{Experimental Details}
For all experiments we set $C_p$, the MCTS parameter balancing exploration vs. exploitation to $\frac{1}{\sqrt{2}}$, as it is a standard value often used in the UCT algorithm~\cite{kocsis2006bandit}. For the selection policy, we always select the branch with the highest score (\Autoref{eq-slct-scoring}), and not randomly as is sometimes done in MCTS. 
All experiments were completed on a Mac Studio 20-core CPU with 64 GB of memory. Experiments were run 10 times, with the average result and standard deviation reported. We experiment 
with different values of the valid rule threshold $\mathcal{V}$, exploration threshold $\theta$ and privacy loss budget $\epsilon$. 
For the baseline protocols, we tested different numbers of queries and 
selected $Q$ as 1000 and 5000 as they provided the best coverage and utility results.  


\subsection{Coverage}\label{sec:exp-cov}


Coverage provides a way to measure how well the learned population ruleset captures the breadth of rule types in the client rulesets. 
We quantify coverage in terms of two metrics: \emph{coverage} and \emph{precision}. In a population ruleset, $R_S$, we define a \emph{valid} rule as one that is contained in the client rule structures of at least $\mathcal{V}$ percent of clients. 
Coverage provides a measure of the number of different rule structures learned by the private model contained in the original client rulesets: 
\begin{equation}
    \text{Coverage} = \frac{|R_{\mathit{valid}}|}{|R_{C_\mathcal{V}}|}
\end{equation}
where $R_{\mathit{valid}}$ is the set of valid rules found in $R_S$ and $R_{C_\mathcal{V}}$ is the set of client rules at the valid rule threshold. 
Precision provides a measure of quality of the overall population ruleset--- Of the rules we found in $R_S$, how many are valid?:
\begin{equation}
    \text{Precision} = \frac{| R_{\mathit{valid}} |}{| R_S |}
\end{equation}

For reasonable privacy loss budgets, it will be impossible to learn all possible client rule structures, but our goal is to learn a set of rules that captures enough of the types of rules contained in the client rulesets to be clinically useful. Coverage, i.e., a wider breadth of rules, is important (as opposed to just learning the top $k$ most common rules,) because the less common rule structures are usually the most informative~\cite{kierner2023taxonomy, berge2023combining}. For example, when clinicians are trying to characterize new conditions or identify new associations indicative of various disease states, typically the less numerous rules characterizing the rare phenomenons are more useful than the more common ones.  In this subsection, we evaluate coverage directly; later, in \Autoref{sec:exp-util}, we evaluate measures of clinical utility.

\shortsection{Comparing Protocols}
\Autoref{fig:cov-vs-base} compares the coverage and precision at different privacy loss budgets for the baseline protocols at 1000 (Baseline~1000Q) and 5000 (Baseline~5000Q) queries compared with the adaptive protocol. We set the valid rule threshold $\mathcal{V}=0.01$ and use $\theta=0.05$ for a controlled comparison and because these metrics align with clinical goals e.g., finding valid rules across 1\% of the population and only allowing a small amount of error at 5\%. 
We experiment with different values of $\mathcal{V}$ and $\theta$ later.  Across all rulesets, the adaptive protocol substantially outperforms the baseline ones. 
Despite having the most rules, the adaptive protocol does the best on the Sepsis ruleset, reaching coverage of 80\% at $\epsilon=1$ with precision above $90\%$ even for the lowest privacy loss budget considered ($\epsilon=0.01$). This is likely because the Sepsis ruleset has the least number of features and many very similar rule structures, requiring less searching through the STL grammar. Contrastingly, the ICU dataset has less patients, resulting in higher noise addition and lower coverage; the T1D ruleset has the most complex rules, with highly disparate rule structures that require deep and wide exploration of the grammar tree, resulting in lower coverage.

\begin{figure*}[t]
     \centering
     \subfigure[ICU]{
         \centering
         \includegraphics[width=0.325\textwidth]{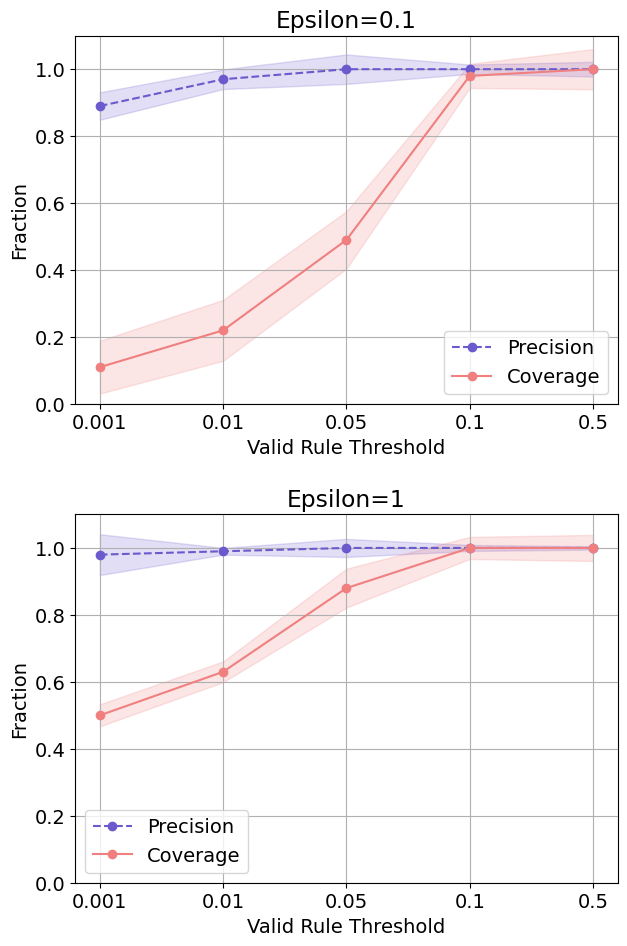}}
     \hfill
     \subfigure[Sepsis]{
         \centering
         \includegraphics[width=0.325\textwidth]{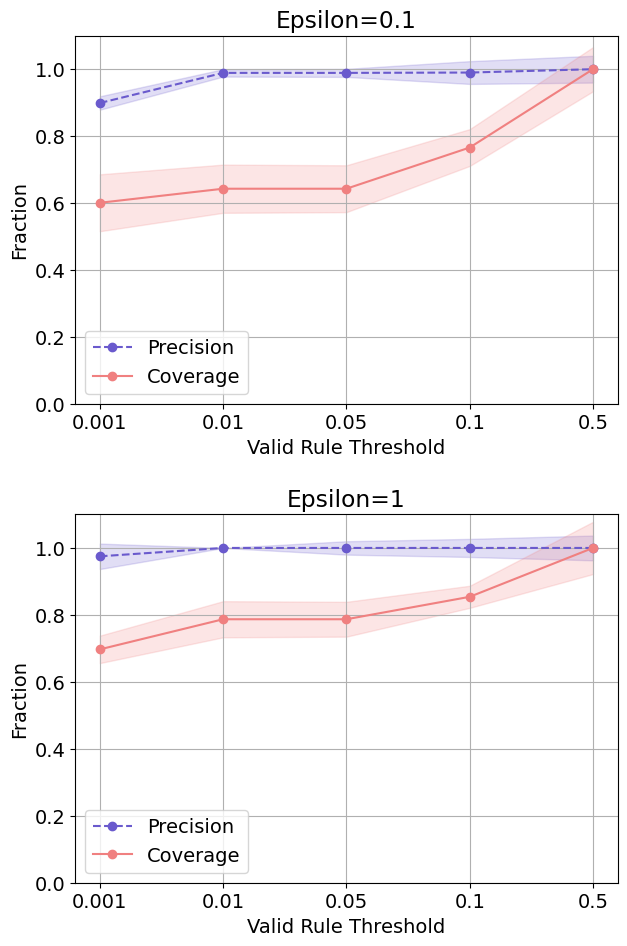}}
    \hfill
    \subfigure[T1D]{
         \centering
         \includegraphics[width=0.325\textwidth]{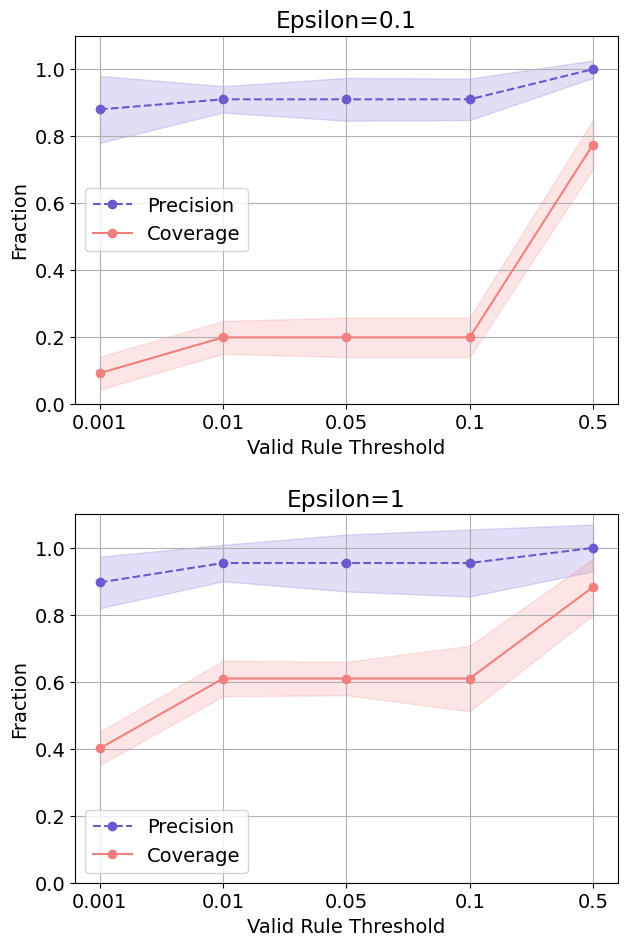}}
    \caption{Coverage at Different Valid Rule Thresholds $\mathcal{V}$ for $\epsilon=0.1$ and $\epsilon=1$}
    \label{fig:cov-compare-Vs}
    \Description{}
\end{figure*}

\shortsection{Query Analysis} \Autoref{fig:cov-query-analysis} compares the privacy loss budget per query across all protocols. As to be expected, the adaptive budget jumps around between different amounts of budget, and the baseline methods use a uniform amount at each query. For the ICU and T1D rulesets, the adaptive budget tends to use higher amounts of budget in the later queries (e.g., after query number $\sim$1700 in T1D), whereas the budget jumps around fairly consistently through all the queries for sepsis.


\shortsection{Comparing $\mathcal{V}$s}
The valid rule threshold $\mathcal{V}$ influences the adaptive protocol's search as a result of the scoring function (\Autoref{eq-slct-scoring}). 
\Autoref{fig:cov-compare-Vs} looks at the effect of $\mathcal{V}$ on the coverage and precision. 
Across all the rulesets, the precision stays pretty stable and the coverage increases as $\mathcal{V}$ increases. This makes sense as there are fewer rules to find as $\mathcal{V}$ increases (see \Autoref{fig:rule-breakdown} which shows the number of ground truth client rules contained at each population percentage). The coverage is decent at lower valid rule thresholds, providing evidence that the adaptive protocol is able to find rare client rules (e.g., rules that are contained by smaller percentages of clients).

\begin{figure*}[t]
     \centering
     \subfigure[ICU]{
         \centering
         \includegraphics[width=0.325\textwidth]{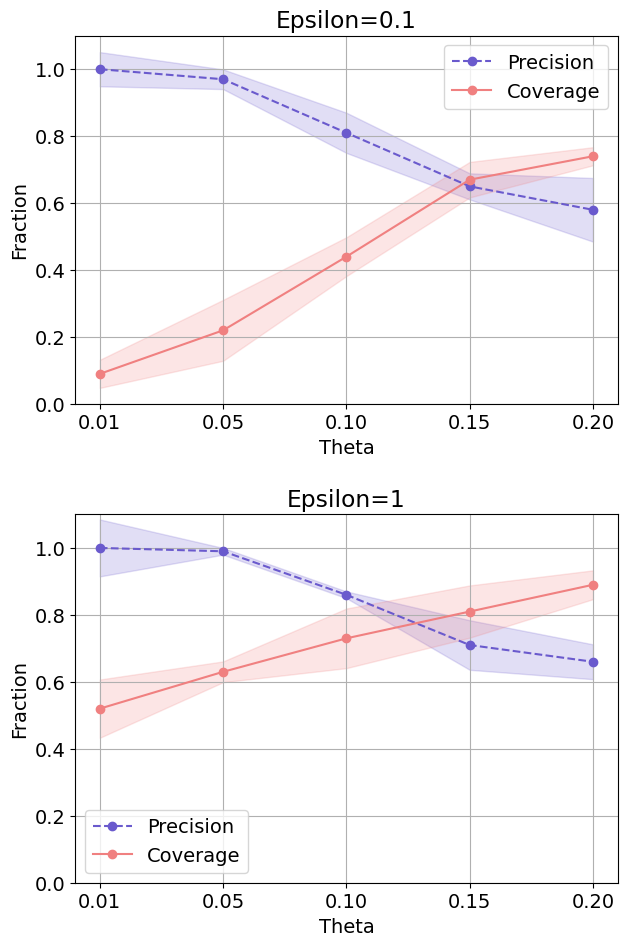}}
     \hfill
     \subfigure[Sepsis]{
         \centering
         \includegraphics[width=0.325\textwidth]{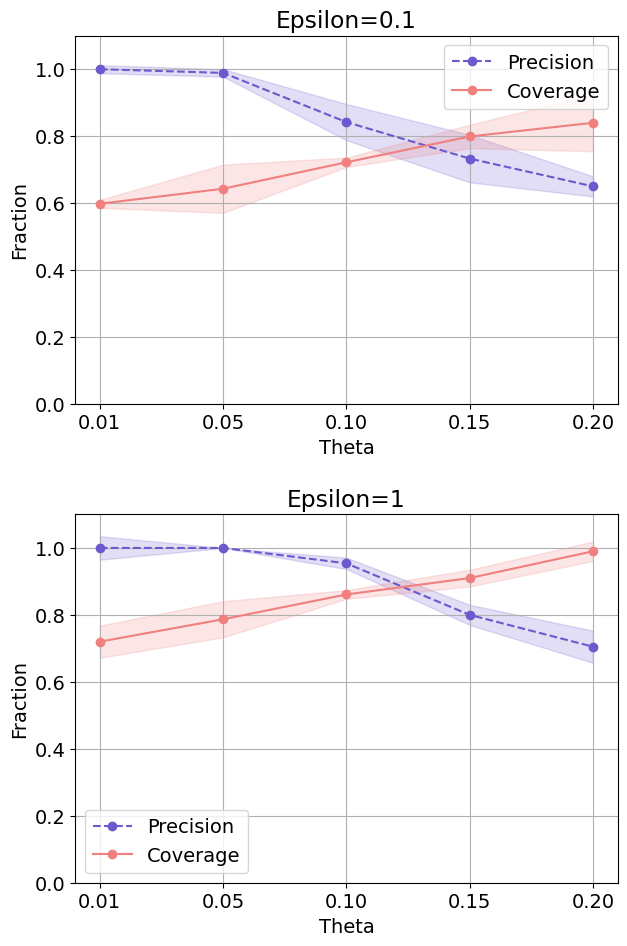}}
    \hfill
    \subfigure[T1D]{
         \centering
         \includegraphics[width=0.325\textwidth]{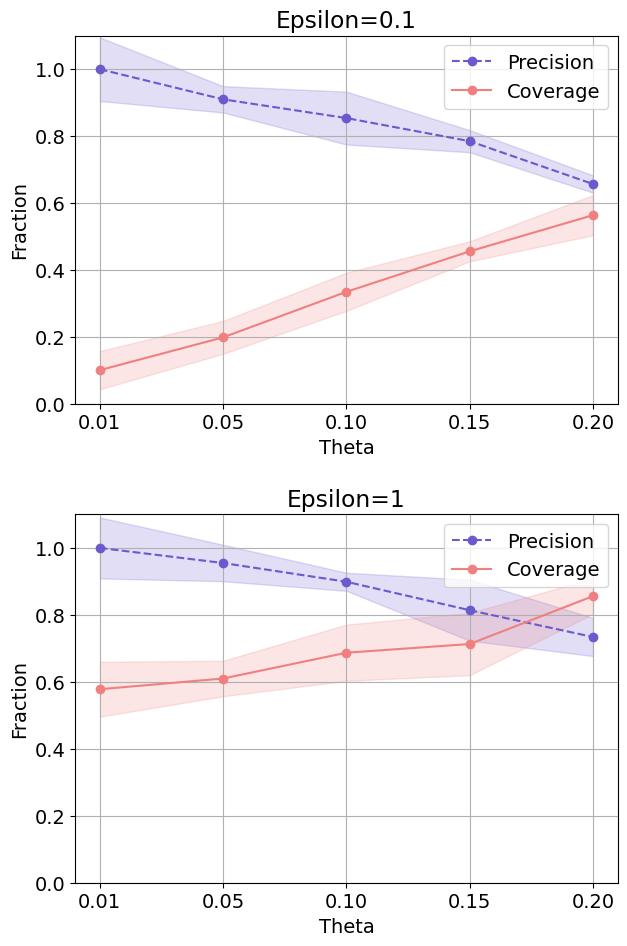}}
    \caption{Effect of Exploration Threshold $\theta$ on Coverage for $\epsilon=0.1$ and $\epsilon=1$}
    \label{fig:cov-compare-thetas}
    \Description{}
\end{figure*}

\shortsection{Impact of Exploration Threshold ($\theta$)}
$\theta$ is the exploration trade-off threshold and determines the probability of falsely ignoring a branch in the adaptive budget allocation. 
\Autoref{fig:cov-compare-thetas} shows the effect of $\theta$ on the coverage results.
There is a trade-off between the precision and coverage dependent on the amount of error allowed. 
Across all rulesets, as $\theta$ increases, the coverage increases, but this comes at a cost to the precision, which drops significantly. This makes sense: as the searching permits more error, more rule structures are found (increasing the coverage) but more invalid rules, rules not actually contained by the client rulesets, are found and returned as a result of the noise in the randomized response querying. Alternatively, lower $\theta$ results in lower coverage, but higher precision. For clinical uses, it is better to favor higher precision since we want very few invalid rules in the learned population ruleset (and we note that this is why a more stringent bound of $\theta=5\%$ were used for the experiments.)


\subsection{Clinical Utility}\label{sec:exp-util}
For the second half of our experimental evaluation, we look at clinical utility, evaluating how useful the rulesets are for representative clinical applications. Since these applications are highly dependent on clinical context, we next describe motivating use cases about how the rulesets may be used and to motivate how utility is evaluated within that context. 

\shortsection{Intensive Care (ICU)} This dataset seeks to understand predictors of clinical deterioration in the ICU. Deterioration refers to a patient's quick onset of a declining physical state that may result in life-threatening outcomes such as death. 
Symptoms of deterioration are highly variable between patients, especially because the condition may occur with little to no warning. 
For our experiments, we evaluate how predictive the learned population rules are at predicting ICU deterioration within the next 15 minutes for each patient. 

\begin{figure*}[t]
     \centering
     \subfigure[ICU]{
         \centering
         \includegraphics[width=0.325\textwidth]{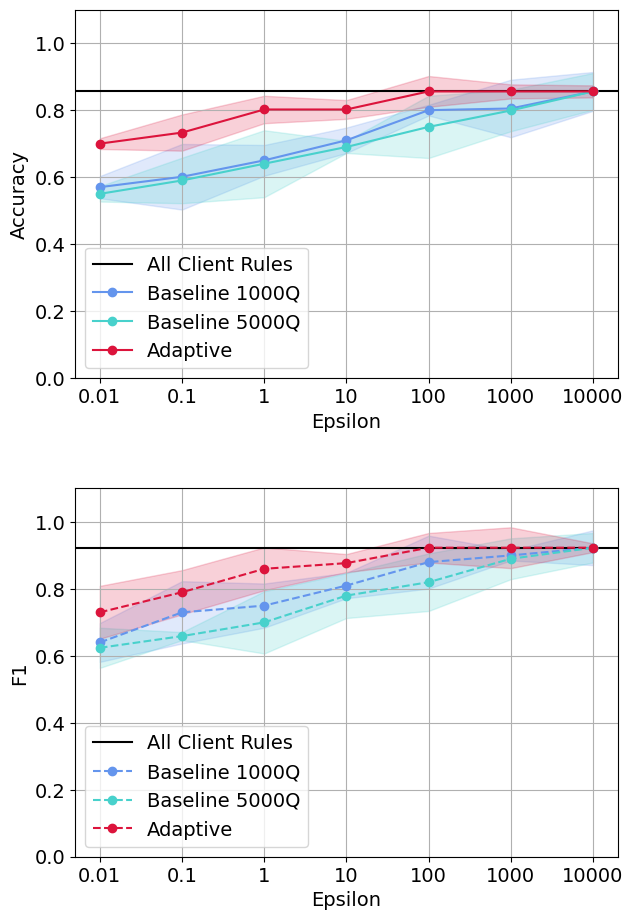}}
     \hfill
     \subfigure[Sepsis]{
         \centering
         \includegraphics[width=0.325\textwidth]{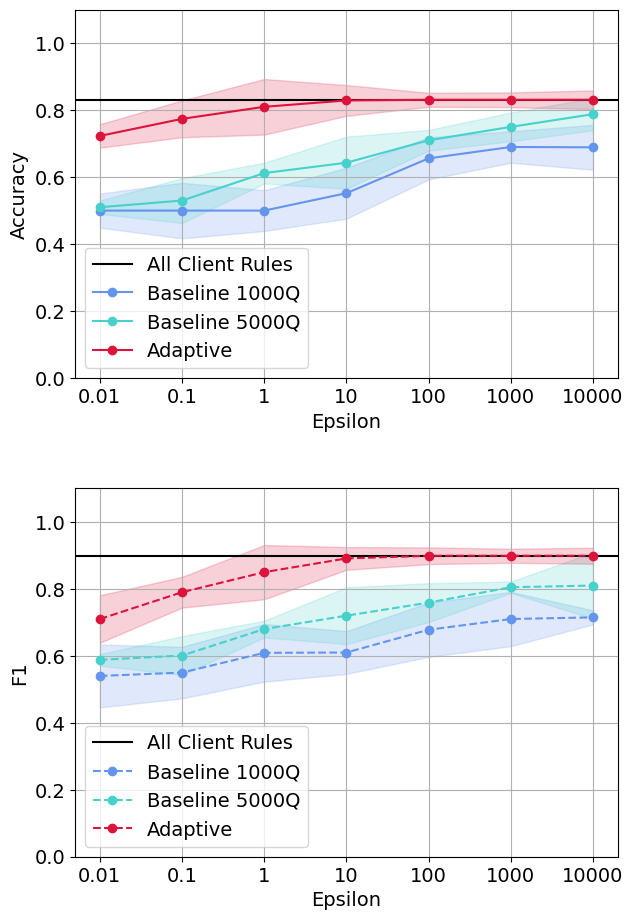}}
    \hfill
    \subfigure[T1D]{
         \centering
         \includegraphics[width=0.325\textwidth]{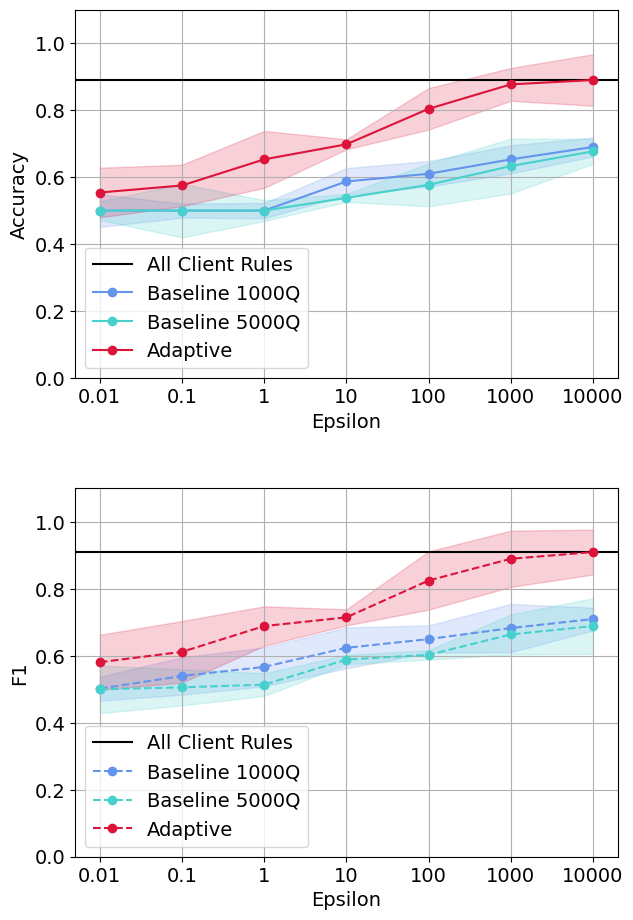}}
    \caption{Clinical Utility ($\mathcal{V}=1\%$, $\theta=5\%$)}
    \label{fig:clin-util}
    \Description{}
\end{figure*}

\begin{figure*}[t]
     \centering
     \subfigure[ICU]{
         \centering
         \includegraphics[width=0.325\textwidth]{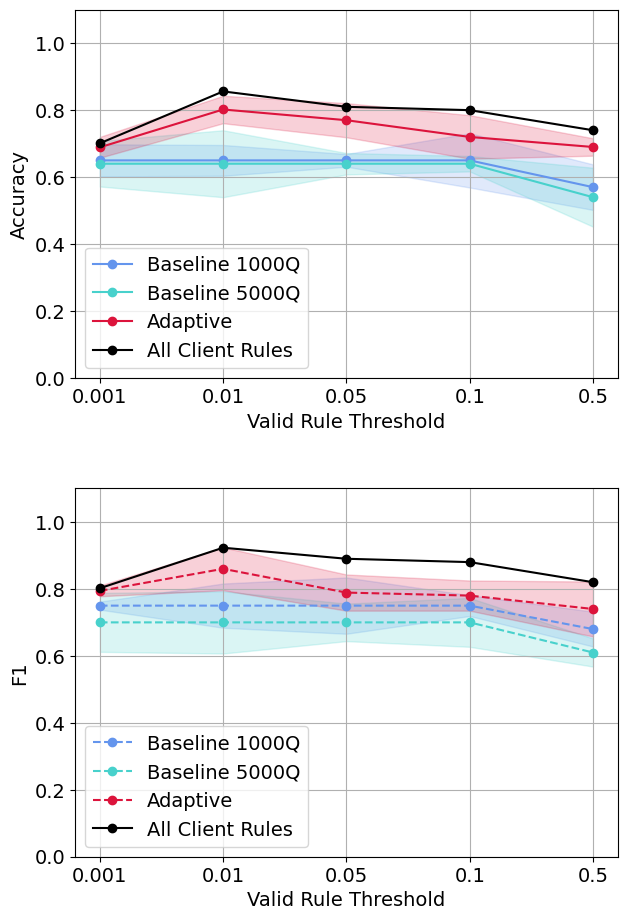}}
     \hfill
     \subfigure[Sepsis]{
         \centering
         \includegraphics[width=0.325\textwidth]{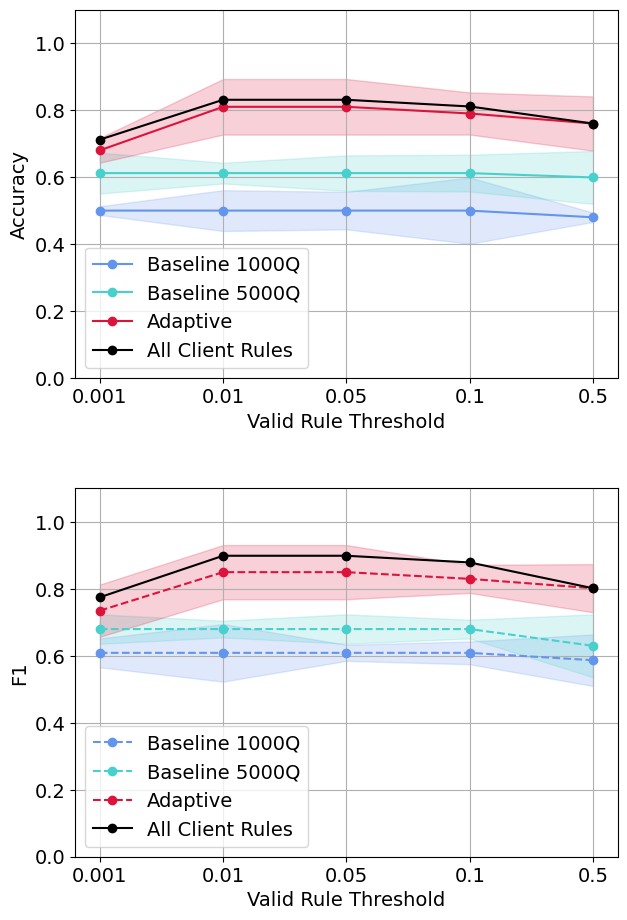}}
    \hfill
    \subfigure[T1D]{
         \centering
         \includegraphics[width=0.325\textwidth]{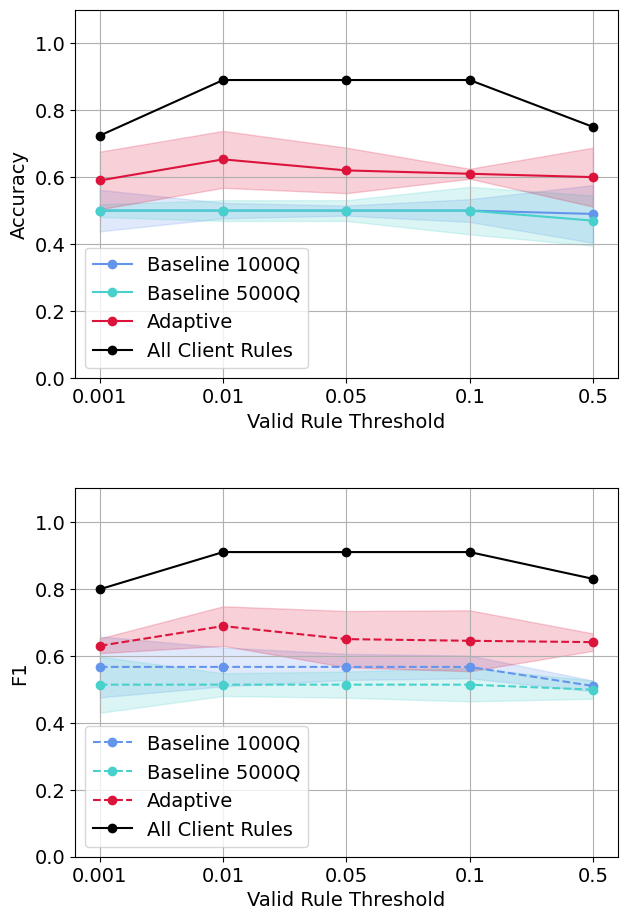}}
    \caption{Comparing Utility Across $\mathcal{V}$s ($\epsilon=1$, $\theta=0.05$ for adaptive protocol)}
    \label{fig:util-compare-Vs}
    \Description{}
\end{figure*}

\shortsection{Sepsis} The Sepsis dataset predicts the onset of sepsis. Sepsis is the body's extreme reaction to an infection it already had and is life-threatening as it can cause a cascade of biological damages such as tissue damage, shock, and organ failure. 
Similar to ICU deterioriation, sepsis occurs extremely quickly, and presents in highly variable ways between patients. In our experiments, and following what a CDSS would be used for, we evaluate how well the learned population rules predict sepsis within the next hour for each patient.

\shortsection{Type I Diabetes (T1D)} The T1D dataset analyzes the glycemic control of individuals with Type I Diabetes. The T1D rules 
characterize the impact of patient behaviors on glycemic outcomes (which dictate better or worse control over the disease). A CDSS might wish to aggregate  subgroup behaviors that characterize good or bad glycemic control. 
In our experiments, we evaluate how predictive the learned rules are for hypoglycemia as one indicator of glycemic control. 

\shortsection{Metrics}
For our use cases, clinical utility indicates the predictive quality of the rules in $R_S$ 
when the rules are used to predict outcomes (deterioration, sepsis and hypoglycemia) on our clinical datasets. Since the rules characterize one of the label classes (the positive or negative class,) the quality of the rule can be judged based on its ability to correctly classify unseen data instances. The rules learned using the privacy-preserving protocols should have predictive quality similar to what would be obtained if the full client rulesets were available. Using a held-out validation data, we use the learned ruleset to predict the binary outcome for each patient using a weighted average taken from each rule in $R_S$, and compute balanced accuracy and F1 based on these predictions. 

\shortsection{Comparing Protocols}
\Autoref{fig:clin-util} displays the utility results of accuracy and F1 scores for the adaptive protocol compared with the two baseline protocols. We set $\mathcal{V}=0.01$ and $\theta=0.05$ for a controlled comparison and because these metrics align with clinical goals e.g., finding valid rules across 1\% of the population and only allowing a small amount of error at 5\%. The black line in each figure displays the ground truth accuracy and F1 for the complete set of client rules at the selected $\mathcal{V}$ together (e.g., all rules from the client ruleset that 1\% of the population have). 
Across all rulesets and privacy loss budgets, the adaptive protocol performs the best.

The ICU and Sepsis datasets have high accuracies and F1s even at lower privacy loss budgets ($\epsilon=0.1$ and $\epsilon=1.0$). However, the T1D dataset performs the worst relative to the other sets, for example with an accuracy of 0.59 for $\epsilon=0.1$. This is likely because T1D individuals have the highest variability in terms of conditions and outcome presentation making it more difficult to correctly predict the outcome hypoglycemia.

\shortsection{Comparing $\mathcal{V}$s}
\Autoref{fig:util-compare-Vs} looks at the effect of the valid rule threshold $\mathcal{V}$ on the clinical utility. All Client Rules refers to the complete set of client rules at the selected $\mathcal{V}$ together. Across all $\mathcal{V}$s, the adaptive protocol outperforms both baseline protocols. 
There is a trade-off between the size of $\mathcal{V}$ and the utility. As $\mathcal{V}$ becomes very small, e.g., $\mathcal{V}=0.001$, accuracy and F1 decrease. This is likely because the number of rules increases substantially as $\mathcal{V}$ decreases, resulting in more varied symptom presentations and increasing disagreement in the rule predictions. In other words, too many unique rules (i.e., rules that only very few patients have) makes it difficult to find a consensus that generalizes across the entire patient cohort. Alternatively, as $\mathcal{V}$ becomes too large, e.g., $\mathcal{V}>0.1$, accuracy and F1 also decrease. This is likely due to the fact that there are few rules contained by large numbers of patients, resulting in only a few general rules that are not as predictive of outcomes. 
As such, there is a sweet spot between the two extremes (which occurs between $\mathcal{V}=0.01$ to $0.05$ for these rulesets), where the population ruleset is generalizable enough to apply to the entire population, but not too general that its predictions are unhelpful.
For all client rules, the highest accuracy and F1 are at $\mathcal{V}=0.01$, which is also why this threshold was used for the prior experiments.

\shortsection{Summary of Findings}
Our experiments demonstrate that the adaptive protocol parameters have a marked effect on the results. Varying $\theta$ results in a trade-off between precision and coverage. As the searching permits more error, more rule structures are found (increasing the coverage) but more invalid rules are also found (decreasing the precision) and vice-versa. Increasing $\mathcal{V}$ results in increased coverage, since there are less total rules to find, but varying $\mathcal{V}$ results in a trade-off for the utility. If $\mathcal{V}$ is too small there are too many rules and the population ruleset does not generalize; if $\mathcal{V}$ is too large, there are too few rules, and the population ruleset is too general to provide nuanced predictions. These findings are useful to help guide real-world instantiations of our protocol. 

Across all experiments, the adaptive protocol outperforms both baseline ones. 
These results are very promising, because they demonstrate that the adaptive protocol is able to learn population rulesets with a breadth of rule types (high coverage) that are clinically useful (high clinical utility), even at low privacy budgets. 
Moreover, the adaptive protocol does well across all three rulesets, despite them having very different characteristics, including different application domains, ruleset sizes, population sizes, rule temporalities and complexity of the rule structures (e.g., length of rules, number of operators/rule.) 
This provides evidence that our protocol may generalize to many different distributed rule-based settings.

%% file: 5.rel-work.tex
\section{Related Work}\label{sec:related}


We discuss relevant LDP prior work, focusing on term collection, tree-based methods and adaptive privacy budgeting. 
We note the clinical rules we collect are different from the kinds of data collected in previous LDP work, 
and no previous work has developed LDP methods for learning logic-based rule structures or for CDSS-specific applications. 
Moreover, no previous methods when applied to the rule-based setting would present a perfect solution. 


\shortsection{Frequent Term Collection}
In the simplest case, one could treat the rules as strings and use prior methods for 
frequent term discovery and collection. 
Prior work in this area has developed LDP models in distributed settings for finding new frequent strings~\cite{Fanti2015}, discovering keystroke data~\cite{kim2018learning, wang2018privtrie}, text mining~\cite{wang2020federated}, frequent item mining~\cite{wang2018locally,wang2019locally,jia2019calibrate} and data mining personal information~\cite{fletcher2019decision}.
These prior methods require large privacy budgets to discover new strings, especially long strings~\cite{wang2020comprehensive}. By taking advantage of the underlying logical structure in our rules (i.e., the rule grammar), we are able to learn long rule structures, even at low privacy budgets. 
Additionally, many of these methods seek to find only the most frequent strings or have poor trade-offs when it comes to finding less frequent strings. For example, \cite{kim2018learning} has high false positive rates for rare unknown words, and \cite{Fanti2015} has low utility for rare n-grams~\cite{wang2020comprehensive, xiong2020comprehensive}. 
We wish to find a \textit{breadth} of rules, as more rare rules tend to be more informative~\cite{kierner2023taxonomy}. By searching a rule grammar and balancing exploration vs. exploitation in our MCTS protocol, we are able to find rare rule structures, and not only the most frequent ones.

\shortsection{Tree-Based LDP Methods}
There has also been prior work in distributed DP protocols that use tree-based methods, either for searching various data spaces or for allocating the privacy loss budget. PrivTrie collects new strings by iteratively building a tree and obtaining a rough estimate of each term prefix by adaptively grouping clients~\cite{wang2018privtrie}. On a related note, LDPART, Zhao et al.\ develop a framework to publish location-record data. They use a hierarchical tree concept (called a partition tree) that extracts relevant location record information and partitions users into groups who are queried to determine whether to keep splitting the sub-nodes or not~\cite{zhao2019ldpart}. 
Our method searches a different data space (rule structures using a grammar) and we do not partition users, which allows them to be queried throughout multiple parts of the tree, and not only the subtree they were partitioned into. This is advantageous because we can use information about the history of previous responses to inform our searching (i.e., in the Backpropagation MCTS step,) and allows our users to be queried in multiple subtrees throughout the exploration tree, resulting in better generalizability of the final population ruleset.


\shortsection{Adaptive Privacy Budgeting}
A straightforward method for adaptive budgeting is to allocate the privacy loss budget using a common scaling factor.
For example, to adaptively allocate the budget at each iteration using an exponential decay mechanism~\cite{ye2019privkv} or at each level in a tree using an increasing geometric or Fibonacci factor~\cite{yan2020arithmetic}. 
Using a uniform scaling strategy as done by these methods is not applicable to our method as there is not a standard factor to guide the scaling (e.g., iteration or tree level). In general, our search dynamically jumps around to different parts of the exploration tree based on the scoring function so it would not make sense to allocate a standard budget amount (e.g., per iteration). Moreover, due to highly complex and varied rule structures, scores are highly variable across tree levels; as such, applying the same budget per level would not be ideal, since many nodes at the same level will have different sensitivities to noise.

Other methods determine the privacy loss budget based on specific algorithm computations, such as halting computations during algorithm runtime~\cite{whitehouse2023fully}, algorithm learning rate for IoT blockchain data~\cite{zhang2022numerical}, ratio of eigenvalues in convolutional neural networks (CNNs) for DP-CNNs~\cite{wangJ2020differential} and tree position and sensitivity for gradient boosted trees~\cite{li2020privacy}.
Although none of these methods are directly applicable to our problem as their computations are derived based on very different domains, they are similar in ideology to our approach: to adjust the privacy loss budget based on an algorithmic computation.

%% file: 6.conclusion.tex
\section{Conclusion}

In this paper, we have presented and evaluated a locally differentally-private framework to learn population rulesets with high coverage and clinical utility for logic-based CDSS. 
This is a first work in a new direction about how to learn complex, structured rules with privacy. Although our work focuses on distributed CDSSs, our protocol can be adapted to fit other distributed settings where aggregating complex rules would be valuable, such as fraud detection and network security monitoring. Moreover, our methodology is amenable to any rule-based learner. Our experimental results demonstrate the promise of learning useful aggregate rulesets across populations while providing strong privacy guarantees.




%% file: appendix.tex
\appendix

\section{Availability}
All code is available at: \url{https://github.com/jozieLamp/DP_Rule_Learning_for_CDSS}

\newpage
\section{Additional MCTS Algorithms}\label{apdx:addtl-algs}

\begin{algorithm}
\caption{SelectNode}
\label{alg-selection}
\SetAlgoLined
\DontPrintSemicolon
\SetNoFillComment
\SetKwProg{Fn}{Function}{:}{end}
\SetKwFunction{selection}{SelectNode}
\Fn{\selection{node $b$, exploration tree $T$, grammar $G$}}{
    \uIf{$b$ is terminal}{\Return $b$\;}
    \uElseIf{any child node of $b$ unvisited}{\selection{unvisited child node}\;}
    \Else{
        \For{all child nodes of $b$ that are not completely explored}{
            $b_{best} \longleftarrow$ child node with the maximum score according to \Autoref{eq-slct-scoring}\;
        }
        \selection{$b_{best}$}
    }
}
\end{algorithm}

\begin{algorithm}
\caption{ExpandNode}
\label{alg-expansion}
\SetAlgoLined
\DontPrintSemicolon
\SetNoFillComment
\SetKwProg{Fn}{Function}{:}{end}
\SetKwFunction{expansion}{ExpandNode}
\Fn{\expansion{node $b_{selected}$, exploration tree $T$, grammar $G$}}{
    \uIf{$b_{selected}$ is terminal or unvisited}{\Return $b_{selected}$\;}
    \Else{
        \If{$b_{selected}$ has no children}{Get all child nodes possible to visit using $G$ and add them to $b_{selected}$}
        \Return {selectionPolicy($b_{selected}$.getChildren())}
    }
}
\end{algorithm}

\begin{algorithm}[h!]
\caption{Backpropagate}
\label{alg-backprop}
\SetAlgoLined
\DontPrintSemicolon
\SetNoFillComment
\SetKwProg{Fn}{Function}{:}{end}
\SetKwFunction{backprop}{Backpropagate}
\Fn{\backprop{node $b$, percentage match count $\hat{c}$, exploration tree $T$}}{
    \While{$b.parent \neq$ None}{
        Add $\hat{c}$ to $b.responses$\;
        $b.visitCount$ += 1\;
        Update $b.score$ according to \Autoref{eq-slct-scoring}\;
        \If{$b$ is terminal or all children of $b$ completely explored}{$b.completelyExplored \longleftarrow$ True\;}
        $b \longleftarrow b.parent$\;
    }
}
\end{algorithm}

\section{Additional Experimental Evaluation}\label{appdx:eval}

\begin{figure}[h]
     \centering
     \subfigure[ICU]{
         \centering
         \includegraphics[width=0.7\linewidth]{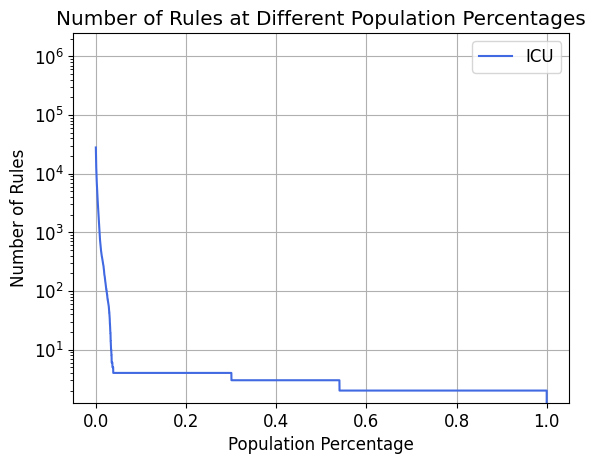}}
     \hfill
     \subfigure[Sepsis]{
         \centering
         \includegraphics[width=0.7\linewidth]{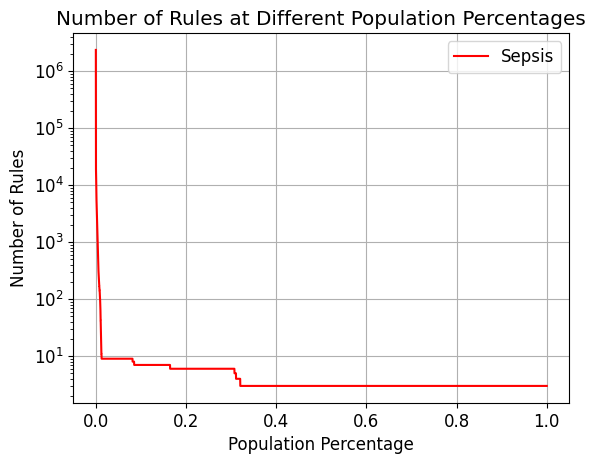}}
    \hfill
    \subfigure[T1D]{
         \centering
         \includegraphics[width=0.7\linewidth]{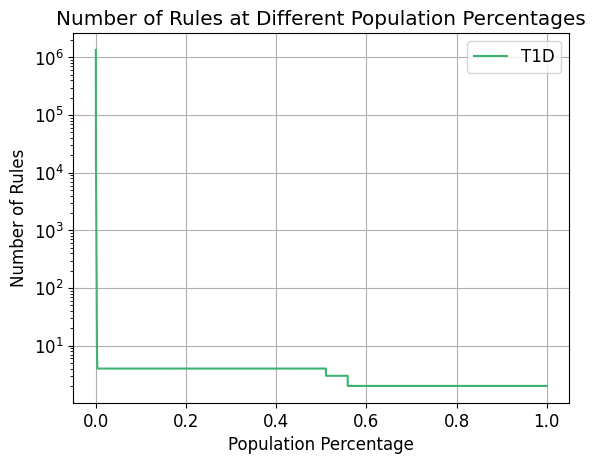}}
    \caption{Rule Breakdown by Population Percentage}
    \label{fig:rule-breakdown}
    \Description{}
\end{figure}
